\documentclass[twocolumn,showpacs,preprintnumbers,amsmath,amssymb,superscriptaddress,floatfix]{revtex4}

\usepackage{graphicx}
\usepackage{bm}

\def\mib#1{\hbox{\boldmath $#1$}}
\def\mbf#1{\hbox{\boldmath $#1$}}
\def\eq#1{Eq.\ (\ref{#1})}

\def\bP{{\mbf P}}
\def\bR{{\mbf R}}

\def\bX{{\mbf X}}

\def\ba{{\mbf a}}
\def\bb{{\mbf b}}

\def\bk{{\mbf k}}

\def\bp{{\mbf p}}
\def\bq{{\mbf q}}
\def\br{{\mbf r}}

\def\bx{{\mbf x}}

\def\bfsigma{{\mbf \sigma}}

\def\SM3{\Sigma N (3/2)}
\def\SN1{\Sigma N (1/2)}

\def\TS1{\hbox{}^3S_1}
\def\TD1{\hbox{}^3D_1}

\def\qf{{\mbf q}_f}
\def\qi{{\mbf q}_i}
\def\abc{(\alpha \beta \gamma)}
\def\H2{1/2}

\begin{document}

\preprint{APS/123-QED}

\title{Faddeev calculation of \mib{3\alpha} and
\mib{\alpha \alpha \Lambda} systems
using \mib{\alpha \alpha} resonating-group method kernel}

\author{Y. Fujiwara}
\affiliation{Department of Physics, Kyoto University, 
Kyoto 606-8502, Japan}%
 \email{fujiwara@ruby.scphys.kyoto-u.ac.jp}

\author{K. Miyagawa}%
\affiliation{Department of Applied Physics,
Okayama Science University, Okayama 700-0005, Japan}

\author{M. Kohno}%
\affiliation{Physics Division, Kyushu Dental College,
Kitakyushu 803-8580, Japan}

\author{Y. Suzuki}%
\affiliation{Department of Physics, Niigata University,
Niigata 950-2181, Japan}

\author{D. Baye}%
\affiliation{Physique Nucl{\'e}aire Th{\'e}orique et Physique
Math{\'e}matique, CP 229, Universit{\'e} Libre de Bruxelles,
B-1050 Brussels, Belgium}

\author{J.-M. Sparenberg}%
\affiliation{TRIUMF, 4004 Wesbrook Mall,
Vancouver, British Columbia, Canada V6T 2A3}

\date{\today}

\begin{abstract}
We carry out Faddeev calculations of three-alpha ($3\alpha$) and
two-alpha plus $\Lambda$ ($\alpha \alpha \Lambda$) systems,
using two-cluster resonating-group method kernels.
The input includes an effective two-nucleon force
for the $\alpha \alpha$ resonating-group method
and a new effective $\Lambda N$ force
for the $\Lambda \alpha$ interaction.
The latter force is a simple two-range Gaussian
potential for each spin-singlet and triplet
state, generated from the phase-shift behavior of the quark-model
hyperon-nucleon interaction, fss2, by using an inversion
method based on supersymmetric quantum mechanics.
Owing to the exact treatment of the Pauli-forbidden states
between two $\alpha$-clusters, the present three-cluster Faddeev formalism
can describe the mutually related, $\alpha \alpha$,
$3\alpha$ and $\alpha \alpha \Lambda$ systems, in terms
of a unique set of the baryon-baryon interactions.
For the three-range Minnesota force which describes
the $\alpha \alpha$ phase shifts quite accurately,
the ground-state and excitation
energies of $\hbox{}^9_\Lambda \hbox{Be}$ are
reproduced within 100 $\sim 200$ keV accuracy.
\end{abstract}

\maketitle



\section{Introduction}
In spite of much effort to incorporate microscopic features
of the alpha-alpha ($\alpha \alpha$) interaction,
a consistent description of the three-alpha ($3\alpha$) and
two-alpha plus $\Lambda$ ($\alpha \alpha \Lambda$) systems
has not yet been obtained so far in the Faddeev formalism.
The most favorable description of the $\alpha \alpha$ system is
the $\alpha \alpha$ resonating-group method (RGM) \cite{SA77}.
Although some of the previous works deal with
the $\alpha \alpha$ RGM kernel explicitly in the $3\alpha$-cluster
Faddeev formalism, they usually yield a large overbinding
for the ground state and sometimes involve spurious states
because of an incomplete treatment of the Pauli-forbidden states
in the $3\alpha$ model space \cite{SC80,KO86,OK89,OK94}.
Various types of $3\alpha$ orthogonality
condition models (OCM) \cite{HO75,HI97,ocm03} also yield
a similar overbinding for the ground state, although
the effect of the Pauli principle between $\alpha$ clusters is
satisfactorily treated in each framework. Only one exception
for this rule is the $3\alpha$ OCM in Refs.~\cite{TB03,DE03}, in which
the Pauli forbidden components described by
the $\alpha \alpha$ bound-state solutions
of the deep Buck, Friedrich,
and Wheatley (BFW) potential \cite{BF77} is
completely eliminated.
The result is rather similar to the
traditional $3\alpha$ Faddeev calculation
using Ali-Bodmer phenomenological $\alpha \alpha$ potential
with a repulsive core \cite{AL66}.
In these calculations, the ground-state
energy of the $3\alpha$ system is less than 1.5 MeV, and
a simultaneous description of the compact shell-model like
ground state and the excited $0^+$ state with well-developed
cluster structure is not possible.
The origin of the different conclusions
in Refs.~\cite{ocm03} and \cite{TB03,DE03}, is spelled out
in Ref.~\cite{AFB},
in which the existence of almost forbidden Faddeev
components inherent to this $3\alpha$ OCM using the bound-state
Pauli-forbidden states of the BFW potential is essential.

A possible resolution of this overbinding problem
of the $3\alpha$ model is found in our new three-cluster
Faddeev formalism, which uses singularity-free $T$-matrices
(RGM $T$-matrices) generated from
the two-cluster RGM kernels \cite{TRGM}.
In this formalism, solving the Faddeev equation automatically
guarantees the elimination of the three-cluster
redundant components from the total wave function.
The explicit energy dependence inherent
in the exchange RGM kernel is self-consistently treated.
We first applied this formalism to
the three-dineutron and $3\alpha$ systems,
and obtained complete agreement between
the Faddeev calculations and variational calculations
using the translationally invariant
harmonic-oscillator (h.o.) basis \cite{TRGM,RED}.
Next, this formalism was applied to a Faddeev
calculation of the three-nucleon bound state \cite{triton},
which employs complete off-shell $T$-matrices derived from the
non-local and energy-dependent RGM kernels of the
quark-model $NN$ interactions, FSS \cite{FSS} and fss2 \cite{fss2}. 
The fss2 model yields a triton binding
energy $B_t=8.519$ MeV in the 50 channel calculation,
when the $np$ interaction is employed
for all the $NN$ pairs in the isospin basis \cite{PANIC02}.
The effect of the charge dependence
of the two-body $NN$ interaction is estimated
to be $-0.19$ MeV for the triton binding energy \cite{MA89}.
This implies that our result is not overbinding in comparison with
the empirical value, ${B_t}^{\rm exp}=8.482$ MeV.
If we attribute the difference, 0.15 MeV, to the effect
of the three-nucleon force, it is by far smaller than the
generally accepted values, $0.5 \sim 1$ MeV \cite{No00},
predicted by many Faddeev calculations employing modern
realistic meson-theoretical $NN$ interactions.
We have further applied this three-cluster Faddeev formalism
to the hypertriton system \cite{hypt},
in which the quark-model hyperon-nucleon ($YN$) interactions
of fss2 yield a reasonable result of the
hypertriton properties similar to the Nijmegen
soft-core potential NSC89 \cite{NSC89}.
Most mathematical details for the Faddeev equation,
employed in this calculation, are given in the present paper. 

Here we apply the present three-cluster Faddeev formalism
to the $\alpha \alpha \Lambda$ model
for $\hbox{}^9_\Lambda \hbox{Be}$.
This hypernucleus plays an important role to study
the $\Lambda N$ interaction in the $p$-shell $\Lambda$-hypernuclei. 
From the early time of the hypernuclear study,
$\hbox{}^9_\Lambda\hbox{Be}$ is
considered to be a prototype of $\alpha$-cluster structure,
in which the two $\alpha$ clusters form a loosely bound subsystem 
by the effect of the extra $\Lambda$ hyperon \cite{Bando}.
Since the $YN$ interaction is usually weaker
than the $NN$ interaction, this system is suitable for studying
a subtle structure change of the two-$\alpha$ system
from $\hbox{}^8 \hbox{Be}$.
In fact, in addition to the $1/2^+$ ground
state \cite{BR75,MA83,DA86,DA91} with
the $\Lambda$-separation energy $B_\Lambda(\hbox{}^9_\Lambda
\hbox{Be})=6.71 \pm 0.04~\hbox{MeV}$ \cite{BA90},
the recent $\gamma$-ray spectroscopy \cite{AK02,TA03} has
revealed the existence of two narrow resonances
in the excited states, which are supposed
to be $5/2^+$ and $3/2^+$ states
generated from the small spin-orbit splitting in the
weak coupling picture of $\hbox{}^8 \hbox{Be}(2^+) \times
\Lambda~(\hbox{spin}~S=1/2)$. From a theoretical point of view,
this is the simplest non-trivial system which requires
the Faddeev formalism with two identical particles,
involving three Pauli-forbidden states
between two clusters. Several model calculations
were already done with various frameworks and two-body potentials. 
Hiyama et al.~\cite{HI97} used the OCM for
the $\alpha \alpha$, $3\alpha$ and $\alpha \alpha\Lambda$ systems
and discussed not only the ground state
of $\hbox{}^9_\Lambda \hbox{Be}$, but also the spin-orbit splitting
of the $5/2^+$ and $3/2^+$ states \cite{HI00}.
They employed simple three-range Gaussian potentials
for the $\Lambda N$ interaction \cite{YA83} based on $G$-matrix
calculations of various Nijmegen
and J{\"u}lich $YN$ one-boson-exchange-potential (OBEP) models.
The $\Lambda \alpha$ potentials are generated
from these $\Lambda N$ effective potentials by the folding procedure
with respect to the $(0s)^4$ h.o.\ wave function
of the $\alpha$ cluster.
They introduced a three-$\alpha$ force
and adjusted the $YN$ parameters to reproduce the binding energies
of the $\hbox{}^{12}\hbox{C}$ and $\hbox{}^9_\Lambda
\hbox{Be}$ ground states. 
Filikhin and Gal \cite{FI02} used the Faddeev and Faddeev-Yakubovsky
formalisms to calculate the $\hbox{}^9_\Lambda \hbox{Be}$ and
$\hbox{}^{10}_{\Lambda \Lambda}\hbox{Be}$ ground states.
They used the Ali-Bodmer $\alpha \alpha$ potential \cite{AL66} and
the so-called Isle potential \cite{KU95} for
the $\Lambda \alpha$ interaction.
They included only $S$-wave in the calculation, and reproduced
the $\hbox{}^9_\Lambda \hbox{Be}$ ground-state energy correctly.
However, if one includes higher partial waves
the Ali-Bodmer $\alpha \alpha$ potential yields overbinding
for $\hbox{}^9_\Lambda \hbox{Be}$ by more than 0.5 MeV.
Oryu et al.~\cite{OR00} carried out a $\alpha \alpha \Lambda$ Faddeev
calculation by using the $\alpha \alpha$ RGM kernel and various
types of $\Lambda \alpha$ potentials in the separable
expansion method. Their energy spectrum
of $\hbox{}^9_\Lambda \hbox{Be}$ is reasonable, but the treatment
of the two-$\alpha$ Pauli principle in the $\alpha \alpha \Lambda$ system
is only approximate. Since they neglected the Coulomb force,
a detailed comparison between their calculated results and
experiment is not possible. Cravo, Fonseca,
and Koike \cite{CR02} performed $\alpha \alpha \Lambda$
and $\alpha \alpha n$ Faddeev calculations
by using many $\alpha \alpha$ and $\Lambda \alpha$
potentials with the Coulomb force included between
the two $\alpha$ particles.
From the comparison of the results for the $\hbox{}^9_\Lambda \hbox{Be}$
and $\hbox{}^9 \hbox{Be}$ systems, they found an interesting sign change
of the quadrupole moments and the magnetic moments for some
excited states. They also pointed out a possibility of negative parity
resonances with $\hbox{}^5_\Lambda \hbox{He}+\alpha$
cluster structure in this threshold region.

Our purpose for the $\alpha \alpha \Lambda$ Faddeev
calculations using $\alpha \alpha$ RGM kernels is threefold.
First, we develop a general three-cluster Faddeev formalism
with two identical clusters, in order to apply it to more
complex three-cluster systems like the hypertriton
interacting via quark-model baryon-baryon interactions.
In the hypertriton system, we have to deal with
the $\Lambda NN$-$\Sigma NN$ coupled-channel system which
involves a Pauli-forbidden state at the quark level
in the $\Lambda N$-$\Sigma N$ subsystem.
Since the baryon-baryon interactions in the quark model
are formulated in the two-cluster RGM formalism,
the present three-cluster formalism is most
appropriate to correlate the baryon-baryon interactions 
with the structure of few-baryon systems.
The second purpose is to make a consistent description of
the $\alpha \alpha$, $3\alpha$ and $\alpha \alpha \Lambda$ systems
using effective $NN$ and $\Lambda N$ interactions. 
This attempt is beyond the scope
of the usual OCM framework and the Faddeev formalism
assuming only inter-cluster potentials.
A comparison of the present $3\alpha$ results with the fully
microscopic $3\alpha$ RGM or GCM \cite{FK77,UE77,DE87} is useful
to examine the approximations involved in the present three-cluster
formalism. The third purpose is to present a tractable
effective $\Lambda N$ force for cluster calculations
of various $p$-shell hypernuclei, which is not purely
phenomenological but derived microscopically from
quark-model baryon-baryon interactions. In particular,
this effective $\Lambda N$ force should be able to reproduce
the correct $\Lambda$-separation energy
of $\hbox{}^5_\Lambda\hbox{He}$;
$B_\Lambda(\hbox{}^5_\Lambda\hbox{He})=3.12 \pm 0.02~\hbox{MeV}$.
Such a $\Lambda \alpha$ interaction is indispensable
for, e.g., a $\Lambda \Lambda \alpha$ Faddeev calculation
using the quark-model $\Lambda \Lambda$ interaction \cite{HE6LL}.
In this paper, we derive an effective $\Lambda N$ force
of two-range Gaussian form from the phase-shift behavior
of the quark-model $YN$ interaction, fss2, by using an inversion
method based on supersymmetric quantum mechanics \cite{SB97}.

This paper is organized as follows.
In the next section, the three-cluster Faddeev formalism
with two identical clusters is given, together with
expressions to calculate the expectation values of the
two-cluster Hamiltonian with respect to the solutions
resulting from the Faddeev equation.
The procedure to calculate
the $\Lambda \alpha$ and $\alpha \alpha$ $T$-matrices is
also discussed, as well as the treatment of
the cut-off Coulomb force employed in this paper.
In the third section, we first briefly discuss the results
of the $3\alpha$ Faddeev calculation, and then those of
the $\alpha \alpha \Lambda$ Faddeev calculation.
The final section is devoted to a summary.
Appendix A gives a brief comment on the rearrangement
factors of three-body systems with two identical particles.
The most general case with explicit spin-isospin degrees
of freedom is discussed. 
In Appendix B, we derive a compact formula
to calculate the $\Lambda \alpha$ Born kernel
for arbitrary types of $\Lambda N$ interactions.
Energies are in MeV and lengths in fm throughout. 

\section{Formulation}

\subsection{Faddeev equation for systems with two identical
clusters}

In order to formulate the Faddeev equation for systems
with two identical particles, we follow the notation of
Refs.~\cite{GL,MI93} as much as possible. The Jacobi-coordinate vectors
are specified by the permutation $\abc$,
which is a {\em cyclic} permutation of (123). For example,
the momentum vectors for the coordinate system $\gamma$ in
the unit of $\hbar$ are
defined by
\begin{eqnarray}
& & \bp_\gamma={m_\beta \bk_\alpha-m_\alpha \bk_\beta
\over m_\alpha+m_\beta}\ , \nonumber \\
& & \bq_\gamma={1 \over M}\left[\left(m_\alpha+m_\beta\right)
\bk_\gamma-m_\gamma \left(\bk_\alpha+\bk_\beta\right)\right]
\ , \nonumber \\
& & \bP=\bk_\alpha+\bk_\beta+\bk_\gamma\ ,
\label{form1}
\end{eqnarray}
where $\bk_\alpha$, $\bk_\beta$, $\bk_\gamma$ are single particle
momenta of particles $\alpha$, $\beta$, $\gamma$ with the
masses $m_\alpha$, $m_\beta$, $m_\gamma$, respectively,
and $M=m_\alpha+m_\beta+m_\gamma$ is the total mass.
Three different sets of the Jacobi coordinates, $(\bp_1, \bq_1)$,
$(\bp_2, \bq_2)$, and $(\bp_3, \bq_3)$,  are
related to each other in the standard relationship for
the rearrangement.
We choose the coordinate system $\gamma=3$ as the standard set
of Jacobi coordinates and assume that
particles 1 and 2 are the two identical particles with a common
mass $m_1=m_2$.
We incorporate the symmetry property for the exchange of
particles 1 and 2 into the Faddeev formalism by assuming
the total wave function $\Psi(\bp, \bq)$ as
\begin{eqnarray}
& & \Psi(\bp_3, \bq_3)=\psi(\bp_3, \bq_3)
\pm \varphi(-\bp_1, \bq_1)+\varphi(\bp_2, \bq_2) \nonumber \\
& & \qquad \hbox{with} \qquad \psi(-\bp_3, \bq_3)
=\pm \psi(\bp_3, \bq_3)\ ,
\label{form3}
\end{eqnarray}
where the upper (lower) sign is applied for identical
bosons (fermions).
The requirement $\Psi(-\bp, \bq)=\pm \Psi(\bp, \bq)$ is satisfied
from this ansaz.

In the application to the  $\alpha \alpha \Lambda$ system,
two $\alpha$ clusters are numbered 1 and 2, and the $\Lambda$ hyperon
is numbered 3. Since the technique to handle the rearrangement
of the Jacobi coordinates in the Faddeev formalism
is well known \cite{GL},
we only give the specification scheme of channels
and the final Faddeev equation after partial-wave decomposition.
We give expressions both in the $LS$-coupling and $jj$-coupling
schemes for later convenience. 
For the $\hbox{}^9_\Lambda \hbox{Be}$ system,
the $\gamma$ channel is specified by $\gamma=3$
with $(\gamma \alpha \beta)=(312)$ in \eq{form1}.
A set of quantum numbers in the $\gamma$-channel is
specified by $\gamma=[(\lambda \ell)L\H2]JJ_z$ in the $LS$-coupling
scheme and $[\lambda (\ell\H2)j]JJ_z$ in the $jj$-coupling scheme
with the angular-spin wave functions
\begin{eqnarray}
& & \langle \widehat{\bp}_3, \widehat{\bq}_3 | \gamma \rangle
=\phi_{\alpha_1} \phi_{\alpha_2}
\nonumber \\
& & \times \left\{ \begin{array}{ll}
[ Y_{(\lambda \ell)L} (\widehat{\bp}_3, \widehat{\bq}_3)
~\xi_{{1 \over 2}}(3) ]_{JJ_z}
& (LS\hbox{-coupling}) \\ [2mm]
[ Y_\lambda (\widehat{\bp}_3) [ Y_\ell(\widehat{\bq}_3)
~\xi_{{1 \over 2}}(3) ]_j ]_{JJ_z}
& (jj\hbox{-coupling})\ . \\
\end{array} \right.
\label{form6}
\end{eqnarray}
Here, $Y_{(\lambda \ell)LL_z} (\widehat{\bp}, \widehat{\bq})
=\left[Y_\lambda (\widehat{\bp}) Y_\ell(\widehat{\bq})
\right]_{LL_z}$, $\xi_{{1 \over 2}}(3)$ is the spin wave function
of $\Lambda$, and $\phi_\alpha$ is the internal wave function
of the $\alpha$ particle. 
Similarly, we define the $\beta$ channel
by $\beta=2$ with $(\beta \gamma \alpha)=(231)$,
and a set of quantum numbers $\beta= [ (\ell_1 \ell_2)L\H2]JJ_z$
($LS$-coupling) and $[ (\ell_1\H2)I\ell_2]JJ_z$ ($jj$-coupling) with
\begin{eqnarray}
& & \langle \widehat{\bp}_2, \widehat{\bq}_2 | \beta \rangle
=\phi_{\alpha_1} \phi_{\alpha_2}
\nonumber \\
& & \times \left\{ \begin{array}{ll}
[ Y_{(\ell_1 \ell_2)L} (\widehat{\bp}_2, \widehat{\bq}_2)
~\xi_{{1 \over 2}}(3) ]_{JJ_z}
& (LS\hbox{-coupling}) \\ [2mm]
[~[ Y_{\ell_1}(\widehat{\bp}_2)~\xi_{{1 \over 2}}(3)]_I
Y_{\ell_2}(\widehat{\bq}_2) ]_{JJ_z}
& (jj\hbox{-coupling})\ . \\
\end{array} \right.
\label{form8}
\end{eqnarray}
The $\alpha$ channel is specified
by $\alpha=1$ with $(\alpha \beta \gamma)=(123)$,
and the quantum numbers similar to
those of the $\beta$ channel.
The partial-wave decomposed Faddeev equation
for the two components $\psi$ and $\varphi$ in \eq{form3} is
given by
\begin{widetext}
\begin{subequations}
\label{form11}
\begin{eqnarray}
& & \psi_\gamma(p, q)=\left[ E-{\hbar^2 \over 4 M_N}\left(p^2
+{8+\zeta \over 4 \zeta}q^2\right) \right]^{-1}
\int^\infty_0 {q^\prime}^2 d q^\prime \int^1_{-1} dx
~\langle p|\widetilde{T}_\lambda
\left(E-{\hbar^2 \over 4 M_N}
{8+\zeta \over 4 \zeta}q^2, \varepsilon_\gamma \right)|p_1 \rangle 
\nonumber \\
& & \times
\sum_\beta {1 \over {p_1}^\lambda}
g_{\gamma \beta}(q, q^\prime;x){1 \over {p_2}^{\ell_1}}
\varphi_\beta(p_2, q^\prime)\ ,\label{form11a} \\
& & \varphi_\beta (p, q)=\left[ E-{\hbar^2 \over 8 M_N}
\left({4+\zeta \over \zeta}p^2
+{8+\zeta \over 4+\zeta}q^2\right) \right]^{-1}
{ 1 \over 2}\int^\infty_0 {q^\prime}^2 d q^\prime
\int^1_{-1} dx
\left\{ \langle p|T_{\ell_1}
\left(E-{\hbar^2 \over 8 M_N} {8+\zeta \over 4+\zeta}q^2\right)
|{p_1}^\prime \rangle \right. \nonumber \\
& & \left. \times
\sum_\gamma {1 \over {p^\prime_1}^{\ell_1}}
g_{\beta \gamma}(q, q^\prime;x)
{1 \over {p^\prime_2}^\lambda} \psi_\gamma(p^\prime_2, q^\prime)
+ \langle p|T_{\ell_1} \left(E-{\hbar^2 \over 8 M_N}
{8+\zeta \over 4+\zeta}q^2 \right)
|\widetilde{p_1} \rangle
\sum_{\beta^\prime} {1 \over \widetilde{p_1}^{\ell_1}}
g_{\beta \beta^\prime}(q, q^\prime;x)
{1 \over \widetilde{p_2}^{\ell^\prime_1}}
\varphi_{\beta^\prime}(\widetilde{p_2}, q^\prime) \right\}
\ ,\nonumber \\
& & \label{form11b}
\end{eqnarray}
where $\zeta=(M_\Lambda/M_N)$ is the mass ratio of $\Lambda$ to
the nucleon and
\begin{eqnarray}
\left\{ \begin{array}{l}
p_1=p\left(q^\prime, {1 \over 2}q; x\right) \\
p_2=p\left(q, {\zeta \over 4+\zeta}q^\prime; x\right) \\
\end{array} \ ,\right. \quad
\left\{ \begin{array}{l}
p^\prime_1=p\left(q^\prime, {\zeta \over 4+\zeta}q; x\right) \\
p^\prime_2=p\left(q, {1 \over 2}q^\prime; x\right) \\
\end{array} \ ,\right. \quad
\left\{ \begin{array}{c}
\widetilde{p}_1=p\left(q^\prime, {4 \over 4+\zeta}q; x\right) \\
\widetilde{p}_2=p\left(q, {4 \over 4+\zeta}q^\prime; x\right) \\
\end{array} \ ,\right.
\label{form11c}
\end{eqnarray}
\end{subequations}
%
with $p(q,q^\prime; x) \equiv \sqrt{q^2+{q^\prime}^2+2qq^\prime x}$.
The $T$-matrices, $\widetilde{T}_\lambda$ and $T_{\ell_1}$, are
discussed in Subsecs. II.D and II.C. 
The rearrangement factors
for the $\psi$ - $\varphi$ or $\varphi$ - $\psi$ cross
terms are given by
%
\begin{eqnarray}
g_{\gamma \beta}(q, q^\prime; x)
=g_{\beta \gamma}(q^\prime, q; x)
=\sum_{\lambda_1+\lambda_2=\lambda}
\sum_{\lambda^\prime_1+\lambda^\prime_2=\ell_1}
q^{\lambda^\prime_1+\lambda_2}{q^\prime}
^{\lambda_1+\lambda^\prime_2}
\left({1 \over 2}\right)^{\lambda_2}
\left({\zeta \over 4+\zeta}\right)^{\lambda^\prime_2}
\sum _k (2k+1)~g^{\lambda_1 \lambda^\prime_1 k}_{\gamma \beta}
~P_k(x)\ ,
\label{form12}
\end{eqnarray}
%
where $P_k(x)$ is the Legendre polynomial of rank $k$.
The reduced rearrangement
factor $g^{\lambda_1 \lambda^\prime_1 k}_{\gamma \beta}$ is
expressed as
\begin{eqnarray}
g^{\lambda_1 \lambda^\prime_1 k}_{\gamma \beta}
=\left\{ \begin{array}{ll}
(-1)^\lambda G^{\lambda_1 \lambda^\prime_1 k L}
_{(\lambda \ell), (\ell_1 \ell_2)} 
& (LS{\rm -coupling}) \\
\sum_{L}(-1)^{I+J+L+\ell_1+1}\widehat{j} \widehat{I}
\left(\widehat{L}\right)^2
\left\{ \begin{array}{ccc}
j   & \ell & {1 \over 2} \\
L   &   J  &  \lambda    \\
\end{array} \right\}
\left\{ \begin{array}{ccc}
J   & L & {1 \over 2} \\
\ell_1   &   I  &  \ell_2 \\
\end{array} \right\}
G^{\lambda_1 \lambda^\prime_1 k L}
_{(\lambda \ell), (\ell_1 \ell_2)}
& (jj{\rm -coupling}) \\
\end{array} \right.\ ,
\label{form13}
\end{eqnarray}
%
with $\widehat{j}=\sqrt{2j+1}$ etc.,
and the spatial angular-momentum
factor $G^{\lambda_1 \lambda^\prime_1 k L}
_{(\lambda \ell),(\ell_1 \ell_2)}$ in Appendix \eq{a17}.
For the $\varphi$ - $\varphi$ type rearrangement,
these factors are given by
\begin{eqnarray}
g_{\beta \beta^\prime}(q, q^\prime; x)
=\sum_{\lambda_1+\lambda_2=\ell_1}
\sum_{\lambda^\prime_1+\lambda^\prime_2=\ell^\prime_1}
q^{\lambda^\prime_1+\lambda_2}{q^\prime}
^{\lambda_1+\lambda^\prime_2}
\left({4 \over 4+\zeta}\right)^{\lambda_2+\lambda^\prime_2}
\sum _k (2k+1)~g^{\lambda_1 \lambda^\prime_1 k}_{\beta \beta^\prime}
~P_k(x)\ ,
\label{form14}
\end{eqnarray}
with
\begin{eqnarray}
g^{\lambda_1 \lambda^\prime_1 k}_{\beta \beta^\prime}
=\left\{ \begin{array}{ll}
(-1)^{\ell_1+\ell^\prime_1}
~G^{\lambda_1 \lambda^\prime_1 k L}_{(\ell_1 \ell_2),
(\ell^\prime_1 \ell^\prime_2)}
& (LS{\rm -coupling}) \\
\sum_{L}(-1)^{I-I^\prime}\widehat{I} \widehat{I^\prime}
\left(\widehat{L}\right)^2
\left\{ \begin{array}{ccc}
J      &   L  & {1 \over 2} \\
\ell_1 &   I  &      \ell_2 \\
\end{array} \right\}
\left\{ \begin{array}{ccc}
J      &   L  & {1 \over 2} \\
\ell^\prime_1 &  I^\prime  & \ell^\prime_2 \\
\end{array} \right\}
G^{\lambda_1 \lambda^\prime_1 k L}_{(\ell_1 \ell_2),
(\ell^\prime_1 \ell^\prime_2)}
& (jj{\rm -coupling}) \\
\end{array} \right.\ .
\label{form15}
\end{eqnarray}
\end{widetext}

\subsection{Calculation of \mib{\varepsilon_\gamma}
and \mib{\varepsilon_\beta}}

In this subsection, we derive some formulas to calculate
expectation values of the two-cluster
Hamiltonians, $h_\gamma+V^{\rm RGM}_\gamma(\varepsilon_\gamma)$
and $h_\beta+V_\beta$, where $h_\gamma$ is
the kinetic-energy operator of the $\gamma$-pair etc.
In the present application, $V^{\rm RGM}_\gamma
(\varepsilon_\gamma)$ is
the $\alpha \alpha$ RGM kernel and $V_\beta$ is
the $\Lambda \alpha$ kernel.
We deal with the energy dependence
of the $\alpha \alpha$ RGM kernel self-consistently by
calculating
\begin{eqnarray}
\varepsilon_\gamma=\langle \Psi|\,h_\gamma
+V^{\rm RGM}_\gamma(\varepsilon_\gamma)\,|\Psi\rangle
\label{self1}
\end{eqnarray}
for the normalized Faddeev solution $\Psi$. 
The potential term of the matrix element in \eq{self1} is most
easily obtained from various matrix elements of the kinetic-energy
operators. Suppose $\Psi$ be a sum of three Faddeev components,
$\Psi=\psi_\alpha+\psi_\beta+\psi_\gamma$. Then the Faddeev
equation $(E-H_0)\psi_\gamma=V_\gamma \Psi$ with $V_\gamma
=V^{\rm RGM}_\gamma(\varepsilon_\gamma)$ and $H_0
=h_\gamma+h_{\bar{\gamma}}$ yields $\langle \Psi |V_\gamma|
\Psi \rangle=\langle \psi_\gamma |E-H_0|\Psi \rangle$.
Thus \eq{self1} becomes
\begin{eqnarray}
\varepsilon_\gamma=E\langle \psi_\gamma|\Psi \rangle
-\langle \psi_\gamma |H_0|\Psi \rangle
+\langle \Psi |h_\gamma|\Psi \rangle\ .
\label{self2}
\end{eqnarray}
We can write a similar equation also for the $\beta$ pair.
We calculate $\varepsilon_\beta$, although the self-consistent
procedure is not necessary for the $\Lambda \alpha$ interaction.
The kinetic energy term $\langle \Psi|h_\beta|\Psi \rangle$ is
obtained from $\langle \Psi|h_\gamma|\Psi \rangle$ as follows.
Using the momentum Jacobi coordinates in \eq{form1}, we can
easily show
\begin{eqnarray}
& & (m_\beta+m_\gamma) h_\alpha+(m_\gamma+m_\alpha) h_\beta
+(m_\alpha+m_\beta) h_\gamma \nonumber \\
& & =M H_0\ .
\label{self3}
\end{eqnarray}
For two identical particles with $m_\alpha=m_\beta$,
this relationship yields
\begin{eqnarray}
h_\alpha+h_\beta={M \over m_\beta+m_\gamma} H_0
-{2m_\beta \over m_\beta+m_\gamma} h_\gamma\ ,
\label{self4}
\end{eqnarray}
and
\begin{eqnarray}
\langle \Psi |h_\beta|\Psi \rangle & = &
{M \over 2(m_\beta+m_\gamma)}
\langle \Psi |H_0|\Psi \rangle
\nonumber \\
& &  -{m_\beta \over m_\beta+m_\gamma}
\langle \Psi |h_\gamma|\Psi \rangle\ .
\label{self5}
\end{eqnarray}
Thus we find, for the $\alpha \alpha \Lambda$ system,
\begin{eqnarray}
\varepsilon_\beta & = & E\langle \psi_\beta|\Psi \rangle
+{8+\zeta \over 2(4+\zeta)}\langle \psi_\gamma |H_0|\Psi
\rangle \nonumber \\
& & +{4 \over 4+\zeta} \left[
\langle \psi_\beta |H_0|\Psi \rangle
-\langle \Psi |h_\gamma|\Psi \rangle \right]\ .
\label{self7}
\end{eqnarray}
We need to calculate the overlap matrix
elements $\langle \psi_\gamma|\Psi \rangle$,
$\langle \psi_\beta|\Psi \rangle$, and
$\langle \psi_\gamma |H_0|\Psi \rangle
=\langle \psi_\gamma |H_0|\psi_\gamma \rangle
+2~\langle \psi_\gamma |H_0|\psi_\beta \rangle$,
$\langle \psi_\beta |H_0|\Psi \rangle
=\langle \psi_\gamma |H_0|\psi_\beta \rangle
+\langle \psi_\beta |H_0|\psi_\beta+\psi_\alpha \rangle$,
$\langle \Psi |h_\gamma|\Psi \rangle
=\langle \psi_\gamma |h_\gamma|\psi_\gamma \rangle
+4~\langle \psi_\gamma |h_\gamma|\psi_\beta \rangle
+2~\langle \psi_\beta |h_\gamma|\psi_\beta+\psi_\alpha \rangle$.
These are calculated from $\psi$ and $\varphi$ by using
the recoupling techniques developed in Appendix A.
The final result is
\begin{widetext}
\begin{subequations}
\label{self9}
\begin{eqnarray}
& & \langle \psi_\gamma |H_0|\Psi \rangle
=\sum_\gamma \int^\infty_0 p^2dp\,q^2dq {\hbar^2 \over 4M_N}
\left(p^2+{8+\zeta \over 4\zeta}q^2\right)
\left[\psi_\gamma(p, q)\right]^2 \nonumber \\
& & +\sum_{\gamma, \beta} \int^\infty_0 q^2dq\,{q^\prime}^2dq^\prime
\int^1_{-1} dx\,\psi_\gamma(p_1, q) {\hbar^2 \over 4M_N}
\left({p_1}^2+{8+\zeta \over 4\zeta}q^2\right)
{1 \over {p_1}^\lambda}
g_{\gamma \beta}(q, q^\prime;x){1 \over {p_2}^{\ell_1}}
\varphi_\beta(p_2, q^\prime)\ ,
\label{self9a} \\
& & \langle \psi_\beta |H_0|\Psi \rangle
=\sum_\beta \int^\infty_0 p^2dp\,q^2dq {\hbar^2 \over 8M_N}
\left({4+\zeta \over \zeta}p^2+{8+\zeta \over 4+\zeta}q^2\right)
\left[\varphi_\beta(p, q)\right]^2 \nonumber \\
& & +{1 \over 2}\sum_{\beta, \beta^\prime} \int^\infty_0 q^2dq
\,{q^\prime}^2dq^\prime
\int^1_{-1} dx\,\varphi_\beta(\widetilde{p_1}, q)
{\hbar^2 \over 8M_N}
\left({4+\zeta \over \zeta}{\widetilde{p_1}}^2
+{8+\zeta \over 4+\zeta}q^2\right)
{1 \over {\widetilde{p_1}}^{\ell_1}}
g_{\beta \beta^\prime}(q, q^\prime;x)
{1 \over {\widetilde{p_2}}^{{\ell_1}^\prime}}
\varphi_{\beta^\prime}(\widetilde{p_2}, q^\prime) \nonumber \\
& & +{1 \over 2} \sum_{\gamma, \beta} \int^\infty_0 q^2dq
\,{q^\prime}^2dq^\prime
\int^1_{-1} dx\,\psi_\gamma(p_1, q) {\hbar^2 \over 4M_N}
\left({p_1}^2+{8+\zeta \over 4\zeta}q^2\right)
{1 \over {p_1}^\lambda}
g_{\gamma \beta}(q, q^\prime;x){1 \over {p_2}^{\ell_1}}
\varphi_\beta(p_2, q^\prime)\ .
\label{self9b}
\end{eqnarray}
\end{subequations}
%
The overlap integrals are obtained
by setting $H_0 \rightarrow 1$.
Furthermore, $\langle \Psi |h_\gamma|\Psi \rangle$ is given by
\begin{eqnarray}
& & \langle \Psi |h_\gamma|\Psi \rangle
=\sum_\gamma \int^\infty_0 p^2dp\,q^2dq {\hbar^2 \over 4M_N} p^2
\left[\psi_\gamma(p, q)\right]^2 \nonumber \\
& & +\sum_{\gamma, \beta} \int^\infty_0 q^2dq\,{q^\prime}^2dq^\prime
\int^1_{-1} dx\,\psi_\gamma(p_1, q) {\hbar^2 \over 2M_N} {p_1}^2
{1 \over {p_1}^\lambda}
g_{\gamma \beta}(q, q^\prime;x){1 \over {p_2}^{\ell_1}}
\varphi_\beta(p_2, q^\prime) \nonumber \\
& & +\sum_{\beta, \beta^\prime} \int^\infty_0 p^2dp
\,q^2dq\,\varphi_\beta(p, q) {\hbar^2 \over 8M_N}
\left\{ \left[p^2+\left({8+\zeta \over 4+\zeta}\right)^2 q^2 \right]
\delta_{\beta, \beta^\prime}+{2(8+\zeta) \over 4+\zeta}
pq f_{\beta \beta^\prime} \right\}
\varphi_{\beta^\prime}(p, q) \nonumber \\
& & +\sum_{\beta, \beta^\prime} \int^\infty_0 q^2dq
\,{q^\prime}^2dq^\prime
\int^1_{-1} dx\,\varphi_\beta(\widetilde{p_1}, q)
{\hbar^2 \over 16M_N}
\left( q^2+{q^\prime}^2-2q q^\prime x \right)
{1 \over {\widetilde{p_1}}^{\ell_1}}
g_{\beta \beta^\prime}(q, q^\prime;x)
{1 \over {\widetilde{p_2}}^{{\ell_1}^\prime}}
\varphi_{\beta^\prime}(\widetilde{p_2}, q^\prime)\ .
\label{self10}
\end{eqnarray}
%
Here, $f_{\beta \beta^\prime}$ is given by
%
\begin{eqnarray}
f_{\beta \beta^\prime} & = & \int d \widehat{\bp}_2
d \widehat{\bq}_2
\langle \beta|\widehat{\bp}_2 \widehat{\bq}_2 \rangle
(\widehat{\bp}_2\cdot \widehat{\bq}_2)
\langle \widehat{\bp}_2 \widehat{\bq}_2 |
\beta^\prime \rangle
= (-1)^{\ell_1+\ell^\prime_2}
\widehat{\ell_1} \widehat{\ell_2}
\langle \ell_1 0 1 0|{\ell_1}^\prime \rangle 
\langle \ell_2 0 1 0|{\ell_2}^\prime \rangle
\nonumber \\
& & \times \left\{ \begin{array}{ll}
(-1)^L \left\{ \begin{array}{ccc}
\ell_1  & \ell_2  & L \\
{\ell_2}^\prime &  {\ell_1}^\prime & 1 \\
\end{array} \right\}
& (LS {\rm -coupling}) \\
\sum_L (-1)^{I-I^\prime+L} \widehat{I} \widehat{I^\prime}
(\widehat{L})^2
\left\{ \begin{array}{ccc}
J      &   L  & {1 \over 2} \\
\ell_1 &   I  &      \ell_2 \\
\end{array} \right\}
\left\{ \begin{array}{ccc}
J      &   L  & {1 \over 2} \\
\ell^\prime_1 &  I^\prime  & \ell^\prime_2 \\
\end{array} \right\}
\left\{ \begin{array}{ccc}
\ell_1  & \ell_2  & L \\
{\ell_2}^\prime &  {\ell_1}^\prime & 1 \\
\end{array} \right\}
& (jj{\rm -coupling}) \\
\end{array} \right. \ .
\label{self11}
\end{eqnarray}
\end{widetext}

\subsection{\mib{\Lambda \alpha} \mib{T}-matrix
and effective \mib{\Lambda N} potentials}

The $\Lambda \alpha$ $T$-matrices are obtained by solving
the Lippmann-Schwinger equation
\begin{eqnarray}
T_\ell(p, p^\prime; E) & = & V_\ell(p, p^\prime)
-{4\pi \over (2\pi)^3}{2 \mu \over \hbar^2}
\int^\infty_0 k^2 dk~V_\ell(p, k) \nonumber \\
& & \times {1 \over \gamma^2+k^2}
T_\ell(k, p^\prime; E)\ ,
\label{la1}
\end{eqnarray}
where $\mu=[4\zeta/(4+\zeta)]M_N$ is
the $\Lambda \alpha$ reduced mass
and $E=-(\hbar^2/2\mu)\gamma^2$ is a negative energy.
The partial-wave components $V_\ell(p, p^\prime)$ for
the $\Lambda \alpha$ Born kernel $V(\bp, \bp^\prime)$ are
defined through
\begin{eqnarray}
V(\bp, \bp^\prime)=4\pi \sum_\ell V_\ell(p, p^\prime)
\sum_m Y_{\ell m}(\widehat{\bp})^* Y_{\ell m}
(\widehat{\bp^\prime})\ ,
\label{la2}
\end{eqnarray}
and the $\langle p|T_\ell(E)|p^\prime \rangle$  in \eq{form11b} is
related to $T_\ell(p, p^\prime; E)$ with an extra
factor $4\pi/(2\pi)^3$.

\begin{figure}[b]
\begin{minipage}[t]{85mm}
\includegraphics[angle=-90,width=82mm]{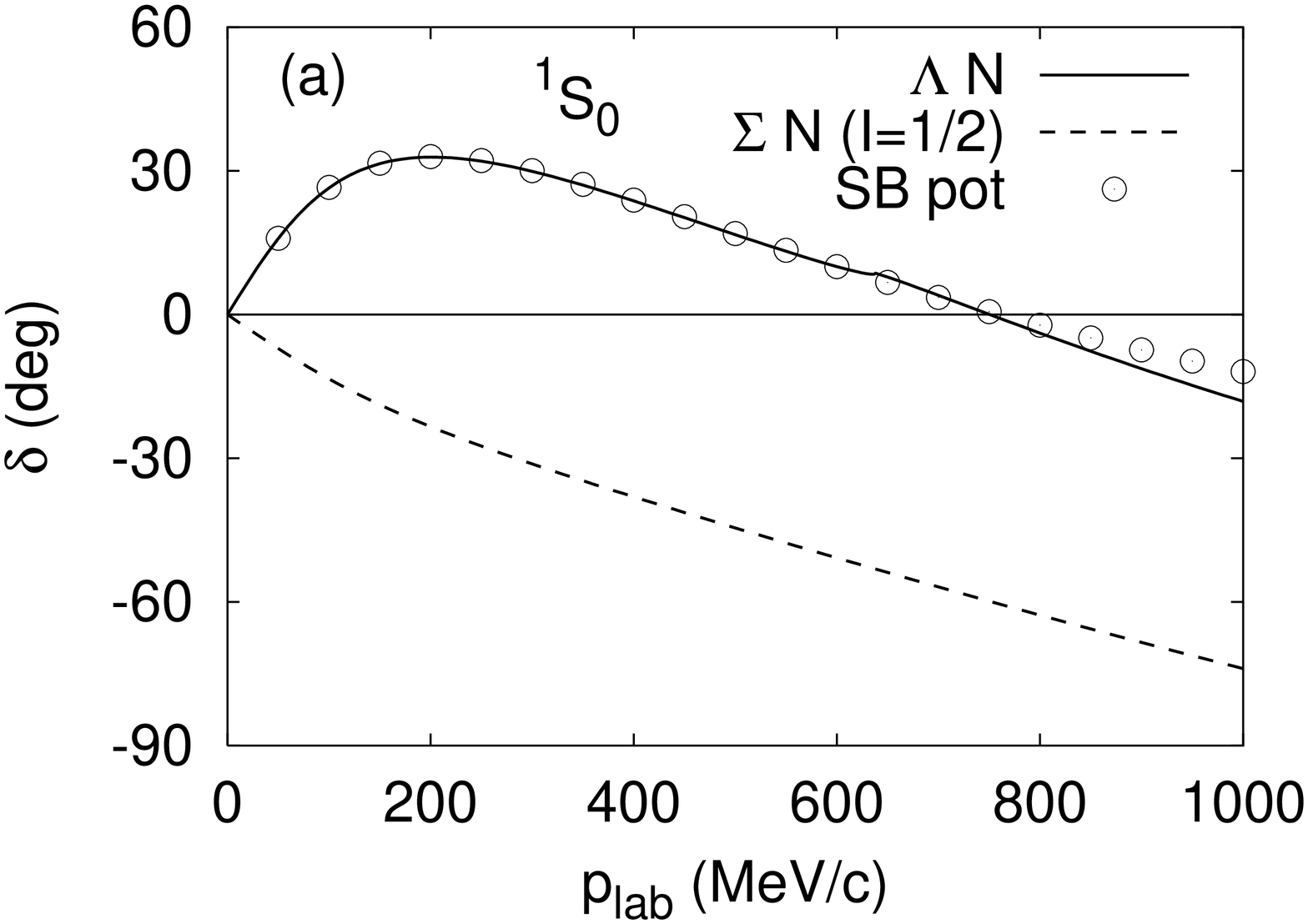}
\includegraphics[angle=-90,width=82mm]{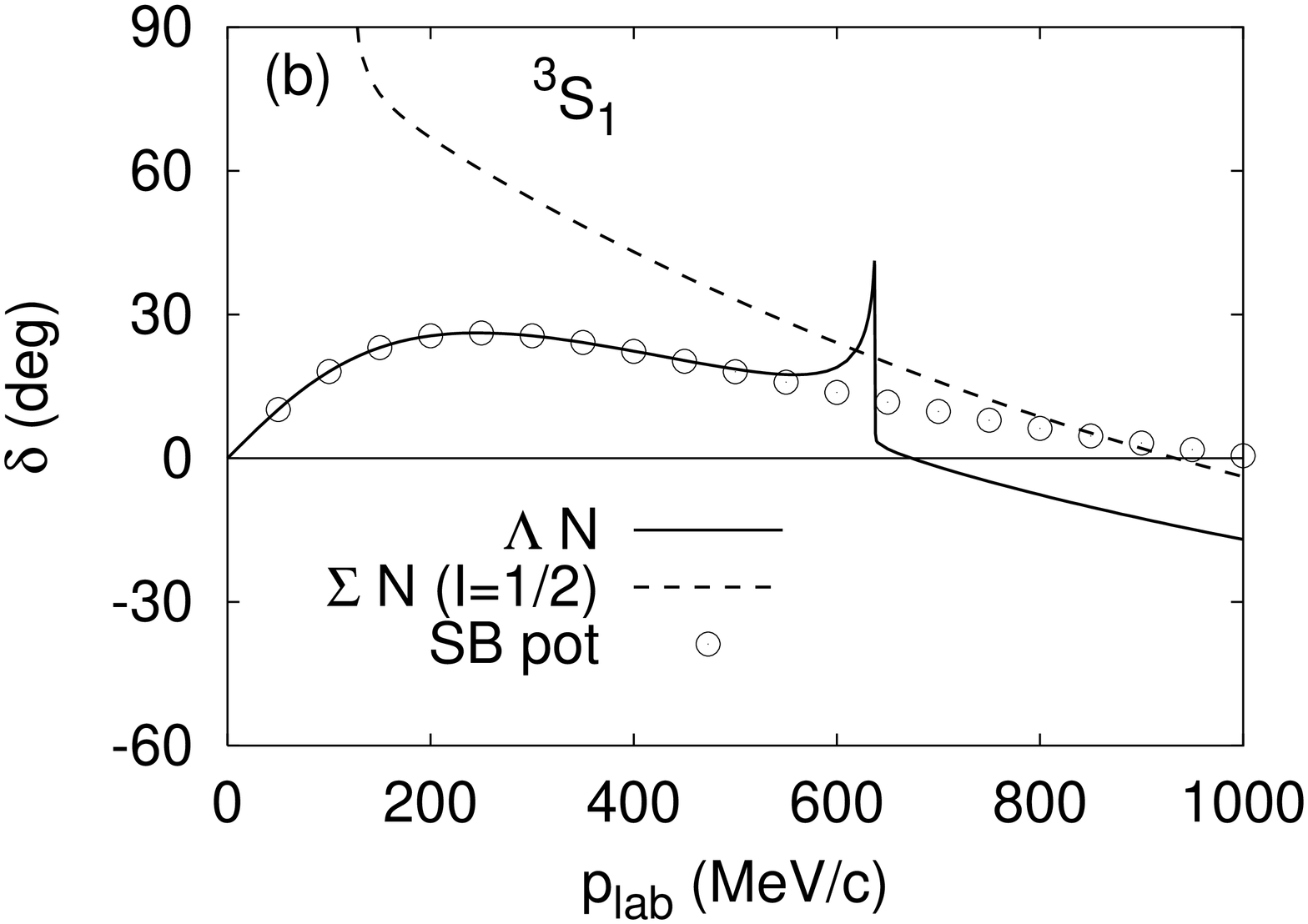}
\caption{$\Lambda N$-$\Sigma N$  $\hbox{}^1S_0$ (a)
and $\hbox{}^3S_1$ (b) phase shifts
for the isospin $I=1/2$ channel,
calculated with fss2 \protect\cite{B8B8} (solid and dashed curves)
and with the SB potential (circles).}
\label{fig1}
\end{minipage}
\end{figure}

For the effective $\Lambda N$ potential, we assume
a Minnesota-type central force \cite{MN77}
\begin{eqnarray}
v_{\Lambda N}& = & \left[\,v(\hbox{}^1E) {1-P_\sigma \over 2}
+v(\hbox{}^3E) {1+P_\sigma \over 2}\,\right]\nonumber \\
& & \times \left[\,{u \over 2}+{2-u \over 2}P_r\,\right]\ ,
\label{la4}
\end{eqnarray}
where $v(\hbox{}^1E)$ and $v(\hbox{}^3E)$ are simple two-range
Gaussian potentials generated from the $\hbox{}^1S_0$ and
$\hbox{}^3S_1$ phase shifts predicted
by the quark-model $\Lambda N$ interaction, fss2.
We use the inversion method based on supersymmetric quantum
mechanics, developed in Ref.~\cite{SB97}, to derive
phase-shift equivalent local potentials. These potentials are
then fitted by two-range Gaussian functions.  
These are given by
\begin{eqnarray}
v(\hbox{}^1S_0) & = & -128.0~\exp(-0.8908~r^2)\nonumber \\
& & +1015~\exp(-5.383~r^2)\ ,\nonumber \\
v(\hbox{}^3S_1) & = & -56.31\,f\,\exp(-0.7517~r^2)\nonumber \\
& & +1072~\exp(-13.74~r^2)\ ,
\label{la5}
\end{eqnarray}
where $f=1$ and $r$ is the relative distance
between $\Lambda$ and $N$.
In the following, we call this
effective $\Lambda N$ potential the SB potential.
Figure \ref{fig1} shows that
these potentials fit the low-energy behavior
of the $\hbox{}^1S_0$ and $\hbox{}^3S_1$ $\Lambda N$ phase shifts
obtained by the full $\Lambda N$-$\Sigma N$ coupled-channel
RGM calculations of fss2.
In the $\hbox{}^3S_1$ state, only the low-energy region is fitted,
since the cusp region cannot be fitted in a single-channel
calculation. This potential overestimates
the $\hbox{}^3S_1$ phase shift in the higher energy region.
The procedure to calculate the $\Lambda \alpha$ Born kernel
for the simple $(0s)^4$ $\alpha$-cluster wave function is 
discussed in Appendix B. Here we only give the final result
for the partial-wave components:
\begin{eqnarray}
V_\ell(q_f, q_i)=\sum^4_{i=1}
\left[X^i_d\,V^d_\ell \left(q_f,q_i;\kappa_i\right)
+X^i_e\,V^e_\ell \left(q_f,q_i;\kappa_i\right)\right].\nonumber \\
\label{la6}
\end{eqnarray}
Here, $X^i_d$ and $X^i_e$ are spin-isospin factors
defined in \eq{b12} and tabulated in Table \ref{table2} for
the present two-range Gaussian potentials. The explicit functional
form of $V^d_\ell \left(q_f, q_i; \kappa_i\right)$ and
$V^e_\ell \left(q_f, q_i; \kappa_i\right)$ are given
in \eq{b13}.

\begin{table}[t]
\caption{$\Lambda \alpha$ spin-flavor coefficients for
the Minnesota-type SB potential with $v=v_0 e^{-\kappa r^2}$. }
\label{table2}
\begin{center}
\renewcommand{\arraystretch}{1.4}
\setlength{\tabcolsep}{3mm}
\begin{tabular}{cccc}
\hline
$i$ & $X^i_d$ & $X^i_e$ & $\kappa_i$ \\
\hline
1,~2   &  ${u \over 2} v_0\left(\hbox{}^1S\right)$
 & $\left(1-{u \over 2}\right) v_0\left(\hbox{}^1S\right)$
 & $\kappa\left(\hbox{}^1S\right)$ \\
3,~4   &  ${3u \over 2} v_0\left(\hbox{}^3S\right)$
 & $3\left(1-{u \over 2}\right) v_0\left(\hbox{}^3S\right)$
 & $\kappa\left(\hbox{}^3S\right)$ \\
\hline
\end{tabular}
\end{center}
\end{table}

In this paper, we also examine the $\Lambda N$ effective
forces \cite{YA83} used by Hiyama et al.\ \cite{HI97} for
comparison. These potentials are generated
from the $G$-matrix calculations of various 
OBEP potentials. They are parameterized as
\begin{eqnarray}
& & v_{\Lambda N}=\sum^3_{i=1}\left\{\left[v^{(i)}_{0~{\rm even}}
+v^{(i)}_{\sigma \sigma~{\rm even}}
\left(\bfsigma_1 \cdot \bfsigma_2\right)\right]
{1+P_r \over 2} \right. \nonumber \\
& & \left. +\left[v^{(i)}_{0~{\rm odd}}
+v^{(i)}_{\sigma \sigma~{\rm odd}}
\left(\bfsigma_1 \cdot \bfsigma_2\right)\right]
{1-P_r \over 2}\right\}~e^{-{\left(r/\beta_i\right)^2}}.
\nonumber \\
\label{la7}
\end{eqnarray}
Since the spin-spin term does not contribute
to the spin saturated $\alpha$-cluster,
the spin-isospin factors in \eq{la6} (with $\kappa_i
\rightarrow 1/(\beta_i)^2$, $i=1$ - 3) are given by
\begin{eqnarray}
X^i_d & = & 2~\left(v^{(i)}_{0~{\rm even}}
+v^{(i)}_{0~{\rm odd}}\right)
\ ,\nonumber \\
X^i_e & = & 2~\left(v^{(i)}_{0~{\rm even}}
-v^{(i)}_{0~{\rm odd}}\right)\ .
\label{la8}
\end{eqnarray}
The explicit values for $v^{(i)}_{0~{\rm even}}$ and
$v^{(i)}_{0~{\rm odd}}$ ($i=1$ - 3) generated from Nijmegen models,
NS, ND, NF, and J{\"u}lich potentials, JA, JB, are given
in Ref.~\cite{HI97}.
[Table V of Ref.~\protect\cite{HI97} includes
a misprint for NS: the width parameters $\beta_i$
for this potential are 1.50 - 1.0 - 0.55,
instead of 1.50 - 0.90 - 0.50 for the other potentials.]

\begin{table}[t]
\caption{Bound-state energies
for the $\Lambda \alpha$ system, $E(\hbox{}^5_\Lambda \hbox{He})$
(in MeV),  calculated by the original SB potential with $f=1$.
The h.o.\ width parameters, 
$\nu=0.275~\hbox{fm}^{-2}$ and $\nu=0.257~\hbox{fm}^{-2}$ are
assumed for the $(0s)^4$ $\alpha$-cluster.
The experimental value is $E^{\rm exp}(\hbox{}^5_\Lambda \hbox{He})
=-3.12 \pm 0.02~\hbox{MeV}$.
}
\label{table3}
\begin{center}
\renewcommand{\arraystretch}{1.2}
\setlength{\tabcolsep}{6mm}
\begin{tabular}{ccc}
\hline
$u$ & $\nu=0.275~\hbox{fm}^{-2}$
& $\nu=0.257~\hbox{fm}^{-2}$ \\
\hline
  1   &  $-4.975$ & $-4.747$ \\
0.6   &  $-4.946$ & $-4.728$ \\
\hline
\end{tabular}
\end{center}
\end{table}

The binding energy of the $\Lambda \alpha$ bound state depends
on the h.o.\ width parameter $\nu$ of the $\alpha$-cluster.
Table \ref{table3} shows that the SB potential of \eq{la5}
overbinds the $\hbox{}^5_\Lambda \hbox{He}$ energy by more than
1.6 MeV. It also shows that the $u$-dependence is very weak, which
implies that $\hbox{}^5_\Lambda \hbox{He}$ is 
an $S$-wave dominated system.
It is well known that a central
single-channel $\Lambda N$ effective
force that fits the low-energy $\Lambda N$ total cross section and
the ground-state energies of $\hbox{}^3_\Lambda\hbox{H}$,
$\hbox{}^4_\Lambda\hbox{H}$ and $\hbox{}^4_\Lambda\hbox{He}$ always
overestimates the $\hbox{}^5_\Lambda \hbox{He}$ binding energy
by more than 2 MeV, due to a lack of $\Lambda$-$\Sigma$ mixing
and the tensor force \cite{DA72,BA80,NE02,KO03}. 
In order to circumvent this difficulty, we introduce a
reduction factor $f$ in the attractive part of
the $\hbox{}^3S_1$ potential in \eq{la5} for
the following Faddeev calculations.
The choices $f=0.8821$ for $\nu=0.275~\hbox{fm}^{-2}$ and
$f=0.8923$ for $\nu=0.257~\hbox{fm}^{-2}$ reproduce the desired
value $E(\hbox{}^5_\Lambda \hbox{He})=-3.120~\hbox{MeV}$,
when the pure Serber type SB potential with $u=1$ is used.
The $\Lambda \alpha$ bound-state energies predicted by
the NS - JB effective $\Lambda N$ potentials
deviate from the original fit in Ref.~\cite{HI97}
by 110 keV - 170 keV ($-3.23$ - $-3.29$ MeV).
This is because they used a slightly different expression
from ours for the exchange term of the $\Lambda \alpha$ potential.
For the $\alpha \alpha \Lambda$ Faddeev calculations
using the Minnesota three-range force for the $\alpha \alpha$ RGM kernel
($\nu=0.257~\hbox{fm}^{-2}$), we readjusted
the strength of the original NS - JB $\Lambda N$ potentials
in order to fit the precise $\hbox{}^5_\Lambda \hbox{He}$
energy, $-3.120$ MeV.
This is achieved by slightly (less than 0.36\%) modifying
the strength of the short-range repulsive term (the third component)
of the original $G$-matrix potentials.

The $\Lambda \alpha$ phase shifts are also calculated, although
there is no experimetal information.
The $S$-wave phase shift shows a monotonic decrease from $180^\circ$,
similar to Fig. 9 of Ref.~\cite{OR00}.
In the energy region $E_{\rm c.m.}(\Lambda \alpha)=0$ - 20 MeV,
the phase shifts of the higher partial waves rapidly 
decrease, starting from 20$\hbox{}^\circ$ - 30$\hbox{}^\circ$ for
the $P$ wave. This implies that the $\Lambda \alpha$ potential
is very much of the Wigner type, and our lack of knowledge
of the $\Lambda \alpha$ interaction
in higher partial waves may not become a serious problem
in the Faddeev calculations.

\subsection{\mib{\alpha \alpha} \mib{T}-matrix
and effective \mib{NN} Potentials}

The $\alpha \alpha$ $T$-matrices used for
the $3\alpha$ and $\alpha \alpha \Lambda$ Faddeev calculations are
generated from the $\alpha \alpha$ RGM kernel
which uses an effective $NN$ potential similar to \eq{la4}.
In the notation used in Ref.~\cite{GRGM},
the $\alpha \alpha$ RGM kernel, $V^{\rm RGM}(\varepsilon)
=V_{\rm D}+{V_{\rm D}}^{\rm CL}+G+G^{\rm CL}
+\varepsilon K$ consists of 
the direct potential $V_{\rm D}$, the direct Coulomb
potential ${V_{\rm D}}^{\rm CL}$, the sum of the exchange
kinetic-energy and interaction kernels, $G=G^{\rm K}+G^{\rm V}$,
the Coulomb exchange kernel $G^{\rm CL}$,
and the exchange normalization kernel $K$.
We have to eliminate redundant components from the
energy-dependent partial wave $T$-matrices,
$T_\lambda(p, p^\prime; E, \varepsilon)$, which satisfy
the Lippmann-Schwinger equation similar to \eq{la1}.
This is necessary only for the $S$-wave ($\lambda=0$) and
$D$-wave ($\lambda=2$) components, for which there exist
two and one h.o.\ Pauli redundant states, $u_{n\lambda}(p)$,
respectively.
Here, $u_{n\lambda}(p)$ are essentially
the h.o.\ wave functions in the momentum representation
with the total h.o.\ quanta $N=2n+\lambda=0$ and 2,
satisfying $Ku_{n\lambda}=u_{n\lambda}$.
They are explicitly given by
\begin{eqnarray}
u_{n\lambda}(p)=(-1)^n {(2\pi)^{3/2} \over \sqrt{4\pi}}
~R_{n\lambda}\left(p,{1 \over 4\mu \nu}\right)
\label{aa0}
\end{eqnarray}
with $\mu=2$, in terms of the standard three-dimensional
h.o.\ radial wave function $R_{n\ell}(r, \nu)$ in
the coordinate representation.
The RGM $T$-matrices defined in Ref.~\cite{TRGM} are
calculated by
\begin{eqnarray}
& & \widetilde{T}_\lambda(p, p^\prime; E, \varepsilon) 
=T_\lambda(p, p^\prime; E, \varepsilon) \nonumber \\
& & +{\hbar^2 \over 4M_N}{\left(\gamma^2+p^2\right)
\left(\gamma^2+{p^\prime}^2\right) \over
\left(\gamma^2+\kappa^2\right)}
\left\{
\begin{array}{l}
\sum^1_{n=0} u_{n0}(p)\,u_{n0}(p^\prime) \\
u_{02}(p)\,u_{02}(p^\prime) \\
\end{array}\right. \nonumber \\
& & \quad \hbox{for} \quad
\lambda=\left\{
\begin{array}{l}
0 \\ 2 \\
\end{array}\right.\ ,
\label{aa1}
\end{eqnarray}
where $\gamma^2=-(4M_NE/\hbar^2)$ and $\kappa^2
=(4M_N\varepsilon/\hbar^2)$.
For the higher partial waves with $\lambda \geq 4$, we define
$\widetilde{T}_\lambda(p, p^\prime; E, \varepsilon)
=T_\lambda(p, p^\prime; E, \varepsilon)$.
The RGM $T$-matrices in \eq{aa1} satisfy the orthogonality
condition
\begin{eqnarray}
& & u_{n\lambda}(p) \nonumber \\
& & ={4 \pi \over (2\pi)^3}{4M_N \over \hbar^2}
\int^\infty_0 {p^\prime}^2 dp^\prime
\widetilde{T}_\lambda(p, p^\prime; E, \varepsilon)
{u_{n\lambda}(p^\prime) \over \gamma^2+{p^\prime}^2}\ ,\nonumber \\
\label{aa2}
\end{eqnarray}
for $n=0,~1$ ($\lambda=0$) and $n=0$ ($\lambda=2$).
Owing to this relationship, we can prove the orthogonality
of the total wave function \eq{form3} to the Pauli-forbidden
states $u_{n\lambda}(p)$.

For the effective $NN$ force, we mainly use the three-range
Minnesota (MN) force \cite{MN77} with
the exchange-mixture parameter, $u=0.94687$, and
the h.o.\ width parameter, $\nu=0.257~\hbox{fm}^{-2}$, for
the $(0s)^4$ $\alpha$-clusters.
We also use the two-range Volkov No.\,1 (VN1)
and No.\,2 (VN2) forces \cite{VO65},
in order to compare our $3\alpha$ results
with the microscopic RGM \cite{FK77,SM04} and GCM \cite{UE77}
calculations. The Majorana parameters $m$ of the Volkov forces
and the h.o.\ width parameters are $m=0.575$ and  
$\nu=0.2515~\hbox{fm}^{-2}$ for VN1, and $m=0.59$ and  
$\nu=0.275~\hbox{fm}^{-2}$ for VN2.
The $\alpha \alpha$ RGM calculations
using these effective $NN$ forces and the complete
Coulomb kernel reasonably reproduce
the empirical $\alpha \alpha$ phase shifts
of the $S$-, $D$-, and $G$-waves,
as well as the $S$-wave resonance
near the $\alpha \alpha$ threshold.
However, the best fit to the
experiment is obtained by the three-range MN force.
For the VN2 force, the $S$-wave resonance appears 
as a bound state with the binding energy $B_{\alpha \alpha}=245$ keV.
Although the VN1 force reproduces this resonance,
the overall fit to the $\alpha \alpha$ phase shifts is less
impressive compared to the MN force.
In the RGM calculation, the precise determination
of the resonance energy is not easy even
in the two-$\alpha$ system, because of the presence of the
Coulomb force. In the present Lippmann-Schwinger formalism
in the momentum representation, the method by Vincent and
Phatak \cite{VP74} is used for solving the scattering problem 
with the full Coulomb force at the nucleon level.
We find that the $0^+$ resonance energy
is 0.18 MeV and 0.14 MeV for VN1 force and the MN force,
respectively. This should be compared with the experimental
value 0.092 MeV.

For the Coulomb force in the $3\alpha$ and $\alpha \alpha
\Lambda$ Faddeev calculations, we use the cut-off Coulomb force
at the nucleon level
\begin{eqnarray}
v^{\rm CL}_{i,j}(r)={1+\tau_z(i) \over 2}{1+\tau_z(j) \over 2}
{e^2 \over r} \cdot \theta(R_C-r)\ ,
\label{aa4}
\end{eqnarray}
with the cut-off radius $R_C$, although an exact treatment
of the point Coulomb force exists for bound-state nuclear
three-body problems with two charged particles \cite{LE84}.
Here $\theta(x)$ is the Heaviside step function.
For the most compact $3\alpha$ ground state, this approximation
with $R_C=10$ fm is good enough to obtain 1 - 2 keV accuracy.  
The Coulomb exchange kernel for \eq{aa4} is calculated analytically.
The partial-wave decomposition of
the $\alpha \alpha$ RGM kernel is carried out numerically using
the Gauss-Legendre 20-point quadrature formula,
when the Coulomb force is not included.
When the cut-off Coulomb force with $R_C=14~\hbox{fm}$ is
employed, it is increased to the 30-point
quadrature formula to obtain an accuracy within 1 keV
for the exchange Coulomb kernel.
The direct Coulomb term is separately integrated
with enough number of numerical integration points.

\section{Results}

To solve the Faddeev equation, we discretize the continuous
momentum variable $p$ ($q$) for the Jacobi coordinate vectors,
using the Gauss-Legendre $n_1$-point ($n_2$-point)
quadrature formula,
for each of the three intervals of~0 - 1 $\hbox{fm}^{-1}$,
1 - 3 $\hbox{fm}^{-1}$ and 3 - 6 $\hbox{fm}^{-1}$. 
The small contribution from the intermediate integral
over $p$ beyond $p_0=6~\hbox{fm}^{-1}$ in
the $\alpha \alpha$ $T$-matrix
calculation is also taken into account by using
the Gauss-Legendre $n_3$-point quadrature formula through the
mapping $p=p_0+{\rm tan}\left\{\pi(1+x)/4\right\}$.
We need $n_1 \geq 10$ and $n_3=5$, so that 35 points are
at least necessary to follow up the inner oscillations
of the two-$\alpha$ bound-state wave function
and the necessary $T$-matrices for solving the Faddeev equation.
These $n_3$ points for $p > 6~\hbox{fm}^{-1}$ are,
however, not included for solving the Faddeev equation,
since it causes a numerical instability for the interpolation.
The momentum region $q=$ 6 $\hbox{fm}^{-1}$ - $\infty$ is
also discretized by the $n_3$ point formula just
as in the $p$ discretization. 
We take $n_1$-$n_2$-$n_3$=15-10-5 for the $3\alpha$ system
and 10-10-5 for the $\alpha \alpha \Lambda$ system, respectively,
unless otherwise specified.
The modified spline interpolation technique developed
in \cite{GL82} is employed to generate
the rearrangement matrices.
For the large-scale diagonalization of non-symmetric matrices,
the Arnoldi-Lanczos algorithm developed in the ARPACK subroutine
package \cite{AR96} is very useful.

\begin{table}[t]
\caption{Results of $3\alpha$ Faddeev calculations, using
the $\alpha \alpha$ RGM kernel, with and without
the Coulomb effect.
The parenthesized numbers indicate the results
when the cut-off Coulomb force
with $R_C=10~{\rm fm}$ are included at the nucleon level.
Partial waves up to $\lambda_{\rm max}$ are included
in $\alpha \alpha$ and $(2\alpha)$-$\alpha$ channels.
The heading $\varepsilon_{2\alpha}$ is
the expectation value of the two-$\alpha$ Hamiltonian
with respect to the 3$\alpha$ bound state solution,
$E_{3\alpha}$ the 3$\alpha$ bound-state energy,
and $c_{(04)}$ the overlap between
the 3$\alpha$ bound-state wave function
and the $SU_3$ (04) shell-model configuration.
For the MN force, the result of the variational calculation
using the translationally invariant h.o.\ basis (h.o.\ var.) is
also given for comparison, where h.o.\ quanta
up to $N=60$ are included.}
\label{table4}
\begin{center}
\renewcommand{\arraystretch}{1.4}
\setlength{\tabcolsep}{0.8mm}
\begin{tabular}{cccccc}
\hline
Force & $\lambda_{\rm max}$ &
$\varepsilon_{2\alpha}$ & $E_{3\alpha}$ & $c_{(04)}$ \\
\hline
     & 4 & 9.657 (10.887) & $-10.751$ ($-5.206$) & 0.900 (0.879) \\
VN1  & 6 & 9.531 (10.779) & $-10.926$ ($-5.365$) & 0.896 (0.875) \\
     & 8 & 9.530 (10.778) & $-10.927$ ($-5.366$) & 0.896 (0.875) \\
\hline
     & 4 & 8.583 (9.608)  & $-11.202$ ($-5.781$) & 0.826 (0.795) \\
VN2  & 6 & 8.449 (9.505)  & $-11.415$ ($-5.967$) & 0.821 (0.790) \\
     & 8 & 8.447 (9.503)  & $-11.417$ ($-5.969$) & 0.821 (0.790) \\
\hline
     & 4 & 12.032 (13.603) & $-15.616$ ($-9.433$) & 0.979 (0.973) \\
MN   & 6 & 11.905 (13.482) & $-15.777$ ($-9.591$) & 0.978 (0.971) \\
     & 8 & 11.904 (13.481) & $-15.779$ ($-9.592$) & 0.978 (0.971) \\
\multicolumn{2}{c}{h.o.\ var.} & 11.903 (13.480)
& $-15.781$ ($-9.594$) & 0.978 (0.971) \\
\hline
\end{tabular}
\end{center}
\end{table}

\subsection{\mib{3\alpha} Faddeev calculation}

In order to make sure that our Faddeev equation is solved
correctly, we first carried out the standard $3\alpha$-particle
Faddeev calculation by using the angular-momentum
dependent Ali-Bodmer potential of $d$ type (ABd).
We find that the $3\alpha$ energy, $E_{3\alpha}=-6.423$ MeV
without Coulomb force, is consistent with previous
calculations \cite{TB03}. Here, we used $\hbar^2/M_\alpha
=10.4465~\hbox{MeV}\cdot\hbox{fm}^2$
and $e^2=1.44~\hbox{MeV}\cdot \hbox{fm}$ for comparison.
When the cut-off Coulomb force is included,
our value $-1.527$ MeV is 4 keV lower than the $-1.523$ MeV given
in Table 1 of Ref.\ \cite{TB03}. This difference is due to
a slightly different treatment of the Coulomb force
between the two calculations.
The small $3\alpha$ binding energy implies that
the Ali-Bodmer phenomenological $\alpha \alpha$ potential
cannot describe the ground state of $\hbox{}^{12}\hbox{C}$ with
a compact shell-model like structure.

On the other hand, the present $3\alpha$ model interacting via
the $\alpha \alpha$ RGM kernel gives enough binding
and a large overlap with the compact shell-model like component.
Table \ref{table4} lists the results of such Faddeev
calculations for the ground state of the $3\alpha$ system
with and without the Coulomb force. The $\alpha \alpha$ RGM
kernels are generated from the VN1, VN2, and MN forces.
When the Coulomb effect is included, the cut-off Coulomb force
with $R_C=10~{\rm fm}$ is employed.
In the last column in Table \ref{table4}, $c_{(04)}$ implies
the overlap amplitude of the $3\alpha$ bound-state function with
the $SU_3$ (04) shell-model configuration.
We find that all three effective $NN$ forces yield
binding energies comparable with the experimental
value $|E^{\rm exp}_{3\alpha}|=7.275$ MeV,
although the result of the $MN$ force is a little too large.
The dominant component of these $3\alpha$ ground states
is the $SU_3$ (04) shell-model configuration.

\begin{table}[b]
\caption{Comparison of the $3\alpha$ ground-state energies,
predicted by the present model ($E_{3\alpha}$) and
by fully microscopic calculations ($E^{\rm full}_{3\alpha}$).
The experimental value is $E^{\rm exp}_{3\alpha}=-7.275$ MeV.
The present model is the Faddeev calculation
using the $\alpha \alpha$ RGM kernel, including the cut-off Coulomb
force with $R_C=10$ fm. 
The heading $E^{\rm int}_\alpha$ implies the internal energy
of the $(0s)^4$ $\alpha$-cluster with the h.o.\ width
parameter $\nu$, $E_{\rm tot}$ the total energy
from the RGM (\protect\cite{SM04} for MN and 
\protect\cite{FK77} for VN2) or
GCM (\protect\cite{UE77} for VN1) calculations,
and $E^{\rm full}_{3\alpha}=E_{\rm tot}-3 E^{\rm int}_{\alpha}$.}
\label{table5}
\begin{center}
\renewcommand{\arraystretch}{1.5}
\setlength{\tabcolsep}{2.2mm}
\begin{tabular}{cccccc}
\hline
Force & $\nu$ ($\hbox{fm}^{-2}$)
& $E^{\rm int}_\alpha$
& $E_{\rm tot}$ & $E^{\rm full}_{3\alpha}$ & $E_{3\alpha}$ \\
\hline
VN1 & 0.2515 & $-27.0$ & $-87.9$ &  $-6.9$ & $-5.37$ \\
VN2 & 0.275  & $-27.3$ & $-89.4$ &  $-7.5$ & $-5.97$ \\
MN  & 0.257  & $-23.9$ & $-83.0$ & $-11.4$ & $-9.59$ \\
\hline
\end{tabular}
\end{center}
\end{table}

In Table \ref{table5} we compare the $3\alpha$ ground-state
energies $E_{3\alpha}$,
predicted in the present three-cluster Faddeev formalism,
with those obtained by fully microscopic calculations,
$E^{\rm full}_{3\alpha}$.
We find that the present three-cluster equation gives $3\alpha$
energies which are only 1.5 - 1.8 MeV higher than those
of the fully microscopic $3\alpha$ RGM or GCM calculations.
This implies that the three-cluster exchange effect,
which is neglected in our three-cluster formalism,
but is present in the fully microscopic three-cluster RGM kernel,
is attractive in nature, and is not as large
as the repulsive three-body force claimed
necessary in the semi-microscopic $3\alpha$ models \cite{SC80,OK89}.
This is mainly because the $3\alpha$ model space used by these
authors does not exclude the $3\alpha$ Pauli-forbidden components
accurately, unlike the one used in the present Faddeev formalism.

\begin{table}[t]
\caption{Kinetic- and potential-energy contributions
to the three-$\alpha$ energy $E_{3\alpha}$,
calculated from $\langle H_0 \rangle
=2\left(3\varepsilon_{2\alpha}-E_{3\alpha}\right)$ and
$\langle V \rangle=3\left(E_{3\alpha}
-2\varepsilon_{2\alpha}\right)$.
The shell-model (04) component, $c_{(04)}$, is large
if $\langle H_0 \rangle$ is large.}
\label{table6}
\begin{center}
\renewcommand{\arraystretch}{1.5}
\setlength{\tabcolsep}{2.2mm}
\begin{tabular}{cccccc}
\hline
Force & $\varepsilon_{2\alpha}$ & $E_{3\alpha}$
& $\langle H_0 \rangle$ & $\langle V \rangle$ & $c_{(04)}$ \\
\hline
VN1 & 10.778 & $-5.366$ &  75.402 &  $-80.768$ & 0.875 \\
VN2 &  9.503 & $-5.969$ &  68.958 &  $-74.927$ & 0.790 \\
MN  & 13.481 & $-9.592$ & 100.068 & $-109.660$ & 0.971 \\
\hline
\end{tabular}
\end{center}
\end{table}

In Tables \ref{table4} and \ref{table5},
we also find that the three-range MN force
gives a somewhat large overbinding of 2 - 4 MeV,
if the $3\alpha$ energy $E_{3\alpha}$ is measured from
the $3\alpha$ threshold.
The decomposition of the $3\alpha$ energy
to the kinetic-energy and potential-energy contributions in
Table \ref{table6} implies that this overbinding is due to the
large cancellation between these two contributions.
In this respect, it is interesting
to note that the $\alpha$ clusters 
with $\nu=0.257~\hbox{fm}^{-2}$ (which gives the correct
rms radius $r_\alpha=(3/4\sqrt{\nu})=1.48~\hbox{fm}$ \cite{KO74}
for the simple $(0s)^4$ $\alpha$-cluster) give
less binding in the framework
of the orthogonality condition model (OCM) \cite{ocm03}.
If the h.o.\ constant parameter $\nu$ is small,
a proper treatment of the $\alpha \alpha$ exchange kernel
seems to be essential in order to obtain a large
binding energy of the $3\alpha$ ground state.
This is reasonable since the large overlap of
two $\alpha$-clusters implies
the importance of nucleon exchange effects.  

\subsection{\mib{\alpha \alpha \Lambda} Faddeev calculation}

For a detailed description of the $\alpha \alpha \Lambda$ bound
states in the Faddeev calculation,
it is important to make sure that the
result is converged with respect to the following three conditions:
\begin{enumerate}
\setlength{\itemsep}{0mm}
\item[1)] convergence with respect to the momentum discretization
points,
\item[2)] convergence with respect to the extension
of partial waves included,
\item[3)] convergence with respect to the cut-off
radius $R_C$, when the cut-off Coulomb force is included.
\end{enumerate}
Among them, the Coulomb effect is the most difficult,
since the $T$-matrix of the full Coulomb force is divergent
at the diagonal part and the strong oscillation
in the momentum representation in the cut-off Coulomb case
does not lead to the correct answer, unless
the numerical angular-momentum projection
of the $\alpha \alpha$ Coulomb kernel (especially the direct
Coulomb term) is accurately performed.
As to the partial waves, we can easily enumerate
all possible angular-momentum states
of $\hbox{}^9_{\Lambda}\hbox{Be}$ for
the $L^\pi=0^+$ ground state with $J=1/2$ and
the $L^\pi=2^+$ excited state
with $J=5/2$ and 3/2 in the $LS$ coupling scheme.
If no $\Lambda \alpha$ spin-orbit force is introduced,
the $J=5/2$ and 3/2 excited states are degenerate
and the $LS$-coupling scheme is more efficient
than the $jj$-coupling scheme
to reduce the number of channels coupled in the calculation.
In the following, the angular-momentum truncation is specified
by $\lambda_{\rm max}$-${\ell_1}_{\rm max}$ values
for the $\alpha \alpha$ and $\Lambda \alpha$ pairs.
For example, $D$-$P$ in the ground-state calculation
implies a 4-channel calculation and $G$-$G$ in
the $L^\pi=2^+$ calculation a 19-channel calculation.
The largest model space adopted is $I$-$I$, which 
is an 11-channel calculation for $L^\pi=0^+$ and a 28-channel
calculation for $L^\pi=2^+$. Note that the variational calculation
in Ref.~\cite{HI97} uses a rather restricted model
space; i.e.\ a three-channel calculation
with $\lambda_{\rm max}=2$ and ${\ell_1}_{\rm max}=0$,
although the meaning of angular-momentum
truncation is a little different from ours. 
For the momentum discretization points,
we find that the energy change due to the increase
of $n_1$-$n_2$-$n_3$ is very much $R_C$ dependent.
It is usually positive if we go from $n_1$-$n_2$-$n_3$=5-5-5
to $n_1$-$n_2$-$n_3$=10-10-5 when the Coulomb force is not included,
but it turns out negative when $R_C=10$ fm and 14 fm.
This implies that the Faddeev calculation
without Coulomb force usually overestimates the binding energy,
if the number of momentum discretization points is not large enough. 
Since the cut-off Coulomb kernels are oscillating,
too small a number of momentum discretization points
such as in $n_1$-$n_2$-$n_3$=5-5-5 case is dangerous
when $R_C$ is very large like $R_C=10$ fm and 14 fm.
The orthogonality to the Pauli-forbidden states also deteriorates
when the number of momentum discretization points is too small.
The squared norm of the Pauli-forbidden components contaminating
the total wave function is typically $10^{-5}$ - $10^{-6}$
when $n_1$-$n_2$-$n_3$=5-5-5, but is improved to
less than $10^{-13}$ for $n_1$-$n_2$-$n_3$=10-10-5.
In this paper, we will mainly show the results
of $n_1$-$n_2$-$n_3$=10-10-5, since the energy gain by further
extension to $n_1$-$n_2$-$n_3$=15-15-5 is usually
less than 1 keV, when the cut-off Coulomb
force with $R_C=10 \sim 14$ fm is included.

The energy gain of the ground state, $\Delta E$, and
that of the self-consistent $\varepsilon_{2\alpha}$ value
by the increase of the maximum angular-momentum values,
$\lambda_{\rm max}$-${\ell_1}_{\rm max}$,
are shown in Table \ref{table7} in the cases when we use
the VN2 or MN forces for the $\alpha \alpha$ interaction
and the SB force for the $\Lambda \alpha$ interaction.
In these calculations the cut-off Coulomb force
with $R_C=6$ fm is employed.
If the $S$-wave calculation is extended to include the $D$-wave,
the energy gain is about 1 MeV for VN2+SB and 1.2 MeV for MN+SB.
The energy gain mainly comes from the partial-wave component
with $\ell_1=\ell_2=1$ of the $\alpha$ - $\hbox{}^5_{\Lambda}
\hbox{He}$ channel. The effect of the partial
wave $\ell_1=\ell_2=2$ is rather small;
i.e., about 50 (VN2) - 60 (MN) keV.
Needless to say, the exact energy gain largely
depends on the character of the $\Lambda N$ odd force.
The ground-state energy is further improved by 7 (VN2) - 5 (MN) keV
and 0.03 (VN1) - 0.0 (MN) keV, according to the extension
to the $G$- and $I$-wave, respectively.
On the other hand, $\varepsilon_{2\alpha}$ is
improved by 160 (VN1) - 281 (MN) keV, 5 - 6 keV and
0.5 - 0.6 keV, according to the extension
to the $D$-, $G$- and $I$-waves, respectively.
In conclusion, partial waves up to the $D$-wave are
sufficient within 10 keV accuracy.
If we wish to have a 1 keV accuracy, we need to take into account
at least up to the $G$-wave.
This implies that the partial-wave truncation
in the Faddeev formalism is very efficient
and the result converges very rapidly, according
to the increase of the partial waves taken into account.
\begin{table}[t]
\caption{
Energy gain for the ground state ($\Delta E$) and
that of the self-consistent $\varepsilon_{2\alpha}$ value
($\Delta \varepsilon_{2\alpha}$) in keV,
for the extension of the maximum angular-momentum values,
$\lambda_{\rm max}$-${\ell_1}_{\rm max}$.
The cut-off Coulomb force with $R_C=6~\hbox{fm}$ is included.
}
\label{table7}
\begin{center}
\renewcommand{\arraystretch}{1.2}
\setlength{\tabcolsep}{2.6mm}
\begin{tabular}{crrrr}
\hline
Force & \multicolumn{4}{c}{VN2+SB} \\
\hline
 & \multicolumn{2}{c}{$\Delta E$~(keV)} &
\multicolumn{2}{c}{$\Delta \varepsilon_{2\alpha}$~(keV)} \\
\hline
$n_1$-$n_2$-$n_3$ & 5-5-5 & 10-10-5 & 5-5-5 & 10-10-5 \\
\hline
S-S $\rightarrow$ D-P & $-954$   & $-954$  & 165  & 160    \\
D-P $\rightarrow$ D-D & $-50$    & $-50$   &   5  &   5    \\
D-D $\rightarrow$ G-G & $-7$     & $-7$    &   6  &   6    \\
G-G $\rightarrow$ I-I & $-0.03$  & $-0.03$ & 0.6  &   0.6 \\
\hline
Force & \multicolumn{4}{c}{MN+SB} \\
\hline
 & \multicolumn{2}{c}{$\Delta E$~(keV)} &
\multicolumn{2}{c}{$\Delta \varepsilon_{2\alpha}$~(keV)} \\
\hline
$n_1$-$n_2$-$n_3$ & 5-5-5 & 10-10-5 & 5-5-5 & 10-10-5 \\
\hline
S-S $\rightarrow$ D-P & $-1165$ & $-1172$ & 287     & 281    \\
D-P $\rightarrow$ D-D & $-57$   & $-58$   &   7     &   7    \\
D-D $\rightarrow$ G-G & $-6$    & $-5$    &   7     &   6    \\
G-G $\rightarrow$ I-I & $-0.1$  & $-0.0$  &   0.5   &   0.5  \\
\hline
\end{tabular}
\end{center}
\end{table}

Table \ref{table8} shows the $R_C$ dependence
of the two-$\alpha$ energy $E(\hbox{}^8\hbox{Be})$,
the self-consistently determined $\varepsilon_{2\alpha}$,
the three-cluster ground-state
energy $E(\hbox{}^9_\Lambda \hbox{Be})$,
the $\Lambda$ separation energy defined
by $B_\Lambda(\hbox{}^9_\Lambda \hbox{Be})=E(\hbox{}^8\hbox{Be})
+M_\Lambda-E(\hbox{}^9_\Lambda \hbox{Be})$, and
the expectation value of the $\Lambda \alpha$ Hamiltonian,
$\varepsilon_{\Lambda \alpha}$,
when the momentum discretization points
with $n_1$-$n_2$-$n_3$=10-10-5 and the partial waves
up to I-I are used in the MN plus SB model.
The energy increase (and the accumulated
one) due to the increase of $R_C$ is also shown with
the plus sign in the second (and the third) row. 
We find that the ground-state
energy $E(\hbox{}^9_\Lambda \hbox{Be})$ increases by 1.621 MeV
when we move from $R_C=0$ to $R_C=6$ fm, which is 
larger than 1.127 MeV calculated
for the free two-$\alpha$ bound state.
This seems to be natural, since the two-$\alpha$ subsystem is more
compact in the $\hbox{}^9_{\Lambda}\hbox{Be}$ system.
The energy increase in the self-consistently
determined $\varepsilon_{2\alpha}$ values is 1.435 MeV,
which is about 200 keV smaller than the energy increase
in $E(\hbox{}^9_\Lambda \hbox{Be})$,
but is still larger than in the free two-$\alpha$ bound state
by about 300 keV.
This observation is a good example that our self-consistent
procedure of determining $\varepsilon_{2\alpha}$ is
reasonably functioning.
It is interesting to note that this large Coulomb effect
in the three-body ground state; i.e., about 1.4 times larger
than in the two-$\alpha$ system, is characteristic
for the increase of $R_C$ from 0 to 6 fm.
For the range from $R_C=6$ to 10 fm, just the opposite is true
and the energy increase in the three-body ground state (85 keV) is
smaller than in the two-$\alpha$ system ($> 133$ keV).
This is apparently because the free $\alpha \alpha$ relative
wave function is more widely spread
than the correlated $\alpha \alpha$ relative wave
function in the $\hbox{}^9_\Lambda \hbox{Be}$ ground state.
The tendency of $\varepsilon_{2\alpha}$ falls just into the middle
of these two extremes.
By using this feature, we can easily estimate the full
Coulomb effect in the $E(\hbox{}^9_\Lambda \hbox{Be})$ ground
state. We find that the result with $R_C=10$ fm is
accurate within a 1 keV error both for $E(\hbox{}^9_\Lambda
\hbox{Be})$ and $\varepsilon_{2\alpha}$.
From Table \ref{table8}, the final result for the MN+SB potentials
is
\begin{eqnarray}
& & E(\hbox{}^9_\Lambda \hbox{Be})
=27.35-34.18=-6.837~\hbox{MeV}\ ,\nonumber \\
& & \varepsilon_{2\alpha}=19.46-18.27=1.181~\hbox{MeV}
\ ,\nonumber \\
& & \varepsilon_{\Lambda \alpha}=9.215-7.954=1.261~\hbox{MeV}
\ ,\nonumber \\
& & c_{(40)}=0.695\ . 
\label{bel1}
\end{eqnarray}
Here we have shown the kinetic-energy and potential-energy
contributions separately in each energy, and $c_{(40)}$ is
the overlap amplitude of the $\hbox{}^9_\Lambda
\hbox{Be}$ ground-state wave function
with the shell-model (40) wave function.
[Note that the sum of the $\varepsilon_{2\alpha}$ potential energy
and twice of the $\varepsilon_{\Lambda \alpha}$ potential energy
is the potential energy of $E(\hbox{}^9_\Lambda \hbox{Be})$,
but this is not true for the kinetic-energy terms.]
We have also carried out the similar analysis
in the VN2+SB model.
The converged result of the VN2+SB forces, including the cut-off
Coulomb force with $R_C=14$ fm, is given by
\begin{eqnarray}
& & E(\hbox{}^9_\Lambda \hbox{Be})
=21.21-28.09=-6.879~\hbox{MeV}\ ,\nonumber \\
& & \varepsilon_{2\alpha}=13.64-12.99=0.649~\hbox{MeV}
\ ,\nonumber \\
& & \varepsilon_{\Lambda \alpha}=8.264-7.548=0.715~\hbox{MeV}
\ ,\nonumber \\
& & c_{(40)}=0.569\ . 
\label{bel2}
\end{eqnarray}
If we compare this result with \eq{bel1} for the MN force,
we find that the energy gain by the more attractive VN2 force
is only 42 keV. 
This result is rather surprising, if we consider that the
VN2 force gives a two-$\alpha$ bound state with 
energy $E_{2\alpha}=-245$ keV. The $\Lambda \alpha$ interaction
by the SB force is also more attractive than in the MN force case
due to the different choice of the h.o.\ width parameter $\nu$.
In other words, the ground state energy
of $\hbox{}^9_\Lambda \hbox{Be}$ is not much affected by
the poor $\alpha \alpha$ and $\Lambda \alpha$ interactions,
as long as we find a well converged value by taking
enough partial waves and a large number of momentum
discretization points. On the other hand, the
$\varepsilon_{2\alpha}$ and $\varepsilon_{\Lambda \alpha}$
values for the MN force are larger than those for the VN2 force
by almost 500 keV. This may be related
to the difference of $\nu$ values in the two calculations.
The smaller $\nu$ value, $0.257~\hbox{fm}^{-2}$,
in the MN force calculation means more extended $\alpha$-clusters
than in the VN2 calculation ($\nu=0.275~\hbox{fm}^{-2}$),
which implies in turn that the relative wave functions
in the $2\alpha$ and $\Lambda \alpha$ subsystems should
be more compact in the MN case. This can be confirmed
by comparing the kinetic-energy contributions
in $E(\hbox{}^9_{\Lambda}\hbox{Be})$, $\varepsilon_{2\alpha}$
and $\varepsilon_{\Lambda \alpha}$ in Eqs.~(\ref{bel1})
and (\ref{bel2}).
For example, the kinetic-energy contribution
in $\varepsilon_{2\alpha}$ is 13.64 MeV in the VN2 case, while
in the MN case it has a much larger value 19.46 MeV.
The compactness of the $\alpha \alpha \Lambda$ relative wave function
in the MN case is also reflected in the fact that $c_{(40)}$ is
larger in the MN case, even though the binding energy is smaller.
Comparing the result in \eq{bel1} with the experimental
value $E^{\rm exp}(\hbox{}^9_{\Lambda}\hbox{Be})
=-6.62 \pm 0.04~\hbox{MeV}$, we can conclude that
the MN+SB combination overbinds
the $\hbox{}^9_{\Lambda}\hbox{Be}$ ground-state energy by 220 keV.  
This is partly because our SB potential is of the pure Serber
type ($u=1$). If we choose $u=0.82$ for the SB force,
the combination with the present MN force
and $\nu=0.257~\hbox{fm}^{-2}$
yields $E(\hbox{}^9_{\Lambda}\hbox{Be})=-6.621~\hbox{MeV}$.
In this case, the $\hbox{}^5_\Lambda\hbox{He}$ bound-state
energy is $-3.105$ MeV.

\begin{table}[t]
\caption{Cut-off radius ($R_C$) dependence of the Coulomb
energies in the two-$\alpha$ bound state
energy $E(\hbox{}^8\hbox{Be})$,
the two-$\alpha$ expectation value $\varepsilon_{2\alpha}$,
the three-body bound state energy $E(\hbox{}^9_\Lambda \hbox{Be})$,
the $\Lambda$ separation energy $B_\Lambda(\hbox{}^9_\Lambda
\hbox{Be})$, and the $\Lambda \alpha$ expectation
value $\varepsilon_{\Lambda \alpha}$.
Calculations are carried out by using $n_1$-$n_2$-$n_3$=10-10-5 and
the partial waves up to I-I.
The three-range MN force and the SB force are used
with $\nu=0.257~\hbox{fm}^{-2}$ for the h.o.\ width
parameter of the $\alpha$-clusters.
The energy increase (and the accumulated one) due to the
increase of $R_C$ is also shown with the plus sign
in the second (and third) row. 
The experimental $\Lambda$ separation energy is 
$B^{\rm exp}_\Lambda(\hbox{}^9_\Lambda \hbox{Be})=6.71 \pm 0.04
~\hbox{MeV}$. The suffix ``ext'' stands for extrapolation.
}
\label{table8}
\begin{center}
\renewcommand{\arraystretch}{1.2}
\setlength{\tabcolsep}{0.6mm}
\begin{tabular}{rrrrrr}
\hline
$R_C$ (fm) & 0 & 6 & 10 & 14 & $\infty$ \\
\hline
$E(\hbox{}^8\hbox{Be})$ & $-1.260$ & $-0.133$ & $>0$ &  & \\
  &  & $+1.127$ & $>+0.133$   &    &  \\
  &  &          & $(>+1.260)$ &    &  \\
$\varepsilon_{2\alpha}$ & $-0.384$ & 1.051 & 1.180 & 1.181
 & $(1.181)_{\rm ext}$ \\
  &  & $+1.435$ & $+0.129$ & $+0.001$   & $-$        \\
  &  &        & $(+1.564)$ & $(+1.565)$ & $(+1.565)_{\rm ext}$  \\
$E(\hbox{}^9_\Lambda \hbox{Be})$ & $-8.543$ & $-6.922$ & $-6.837$
 & $-6.837$ & $(-6.837)_{\rm ext}$ \\
 &   & $+1.621$ & $+0.085$ & $+0.000$   & $-$        \\
 &   &        & $(+1.706)$ & $(+1.706)$ & $(+1.706)_{\rm ext}$ \\
$B_\Lambda(\hbox{}^9_\Lambda \hbox{Be})$
 &  7.283  & 6.789  & $>6.837$ & & \\
\hline
$\varepsilon_{\Lambda \alpha}$ & $1.390$ & $1.228$ & 1.260 & 1.261
&  $(+1.261)_{\rm ext}$ \\
\hline
\end{tabular}
\end{center}
\end{table}

We list the results of various $\Lambda N$ effective
forces used by Hiyama et al.\ in Table \ref{table9},
when they are used in combination with the MN force
for the $\alpha \alpha$ RGM kernel.
The calculations are carried out with $n_1$-$n_2$-$n_3$=10-10-5,
$R_C=10$ fm, and the partial waves up to the $G$-wave,
to obtain the converged results with the accuracy of 1 - 2 keV.

\begin{table}[t]
\caption{$\alpha \alpha \Lambda$ Faddeev calculations
for the $L^\pi=0^+$ ground state, including the cut-off
Coulomb force with $R_C=10$ fm.
The $\alpha \alpha$ RGM kernel is generated
from the three-range MN force with $u=0.94687$ and
$\nu=0.257~\hbox{fm}^{-2}$ for the h.o.\ width
parameter of the $\alpha$-clusters.
The $G$-matrix based effective $\Lambda N$ forces
in Ref.~\protect\cite{HI97} are used
for the $\Lambda \alpha$ interaction, by slightly modifying
the short-range repulsive part to fit the $\Lambda$ separation
energy $B_\Lambda(\hbox{}^5_\Lambda \hbox{He})=3.120~\hbox{MeV}$.
Partial waves up to $\lambda_{\rm max}$ are included
in $\alpha \alpha$-$\Lambda$ channel
and those up to ${\ell_1}_{\rm max}$ are included
in the $\Lambda \alpha$-$\alpha$ channel.
The heading $E(\hbox{}^9_\Lambda \hbox{Be})$ is
the three-body ground-state energy
of $\hbox{}^9_{\Lambda}\hbox{Be}$ in
the $\alpha \alpha \Lambda$ model,
$\varepsilon_{2\alpha}$ the two-$\alpha$ expectation value
determined self-consistently, and $\varepsilon_{\Lambda \alpha}$
the $\Lambda \alpha$ expectation value, and $c_{(40)}$ is
the overlap with the shell model (40) wave function.
}
\label{table9}
\begin{center}
\renewcommand{\arraystretch}{1.0}
\setlength{\tabcolsep}{2.5mm}
\begin{tabular}{ccrrrr}
\hline
force & $\lambda_{\rm max}$-${\ell_1}_{\rm max}$ &
$E(\hbox{}^9_\Lambda \hbox{Be})$
 & $\varepsilon_{2\alpha}$ & $\varepsilon_{\Lambda \alpha}$
 & $c_{(40)}$ \\
\hline
   & S-S & $-5.580$ & 0.909 &  1.136  &   0.606 \\
NS & D-P & $-6.681$ & 1.122 &  1.250  &   0.683 \\
   & D-D & $-6.736$ & 1.133 &  1.255  &   0.686 \\
   & G-G & $-6.743$ & 1.132 &  1.257  &   0.686 \\
\hline
   & S-S & $-5.734$ & 0.764 &  0.774  &   0.579 \\
ND & D-P & $-7.375$ & 1.136 &  0.838  &   0.693 \\
   & D-D & $-7.478$ & 1.159 &  0.842  &   0.697 \\
   & G-G & $-7.483$ & 1.157 &  0.843  &   0.697 \\
\hline
   & S-S & $-5.682$ & 0.802 &  0.882  &   0.587 \\
NF & D-P & $-6.839$ & 1.009 &  0.942  &   0.666 \\
   & D-D & $-6.901$ & 1.021 &  0.944  &   0.669 \\
   & G-G & $-6.906$ & 1.020 &  0.944  &   0.669 \\
\hline
   & S-S & $-5.620$ & 0.862 &  1.030  &   0.599 \\
JA & D-P & $-6.622$ & 1.022 &  1.112  &   0.667 \\
   & D-D & $-6.672$ & 1.031 &  1.114  &   0.669 \\
   & G-G & $-6.677$ & 1.031 &  1.115  &   0.669 \\
\hline
   & S-S & $-5.566$ & 0.915 &  1.154  &   0.606 \\
JB & D-P & $-6.431$ & 1.027 &  1.253  &   0.664 \\
   & D-D & $-6.469$ & 1.034 &  1.255  &   0.666 \\
   & G-G & $-6.475$ & 1.033 &  1.256  &   0.666 \\
\hline
\end{tabular}
\end{center}
\end{table}

Table \ref{table10} lists $\alpha \alpha \Lambda$ Faddeev calculations
for the $2^+$ excited state, including the cut-off
Coulomb force with $R_C=14~\hbox{fm}$.
The momentum discretization points with $n_1$-$n_2$-$n_3$=10-10-5
are employed. When the partial waves are restricted
to D-S or S-D, the $2^+$-state energy is located above
the $\alpha+\hbox{}^5_{\Lambda}\hbox{He}$ threshold with
the threshold energy $-3.12$ MeV.
The listing therefore starts from the 7-channel
calculation with D-P.  
We find that the result is almost converged
with I-I and $R_C=14$ fm, within the accuracy of 1 keV.
The final result for the $2^+$ excited state in the MN+SB model
is
\begin{eqnarray}
& & E=29.47-33.40=-3.926~\hbox{MeV}\ ,\nonumber \\
& & \varepsilon_{2\alpha}=21.55-17.54=4.013~\hbox{MeV}
\ ,\nonumber \\
& & \varepsilon_{\Lambda \alpha}=9.481-7.930=1.551~\hbox{MeV}
\ ,\nonumber \\
& & c_{(40)}=0.645\ . 
\label{bel4}
\end{eqnarray}
If we compare Eqs.~(\ref{bel1}) and (\ref{bel4}),
we find that the 3 MeV excitation energy of the $2^+$ state
mainly comes from an increase of the two-$\alpha$ kinetic
energy (2 MeV) and from the two-$\alpha$ potential energy (1 MeV).
This clearly shows the rotational nature
of the ground $0^+$ and excited $2^+$ states,
composed of the two-$\alpha$ cluster structure
with a weakly coupled $\Lambda$. 
\begin{table}[t]
\caption{Same as Table \protect\ref{table9},
but for the $L^\pi=2^+$ excited state with $R_C=14~\hbox{fm}$.
}
\label{table10}
\begin{center}
\renewcommand{\arraystretch}{1.0}
\setlength{\tabcolsep}{2.5mm}
\begin{tabular}{ccrrrr}
\hline
force & $\lambda_{\rm max}$-${\ell_1}_{\rm max}$ &
$E(\hbox{}^9_\Lambda \hbox{Be})$
 & $\varepsilon_{2\alpha}$ & $\varepsilon_{\Lambda \alpha}$
 & $c_{(40)}$ \\
\hline
   & D-P & $-3.797$ & 3.987 &  1.528  &   0.643 \\
SB & D-D & $-3.874$ & 4.014 &  1.536  &   0.645 \\
   & G-G & $-3.926$ & 4.013 &  1.550  &   0.645 \\
   & I-I & $-3.926$ & 4.013 &  1.551  &   0.645 \\
\hline
   & D-P & $-3.700$ & 3.920 &  1.518  &   0.639 \\
NS & D-D & $-3.772$ & 3.946 &  1.525  &   0.641 \\
   & G-G & $-3.831$ & 3.942 &  1.544  &   0.641 \\
   & I-I & $-3.831$ & 3.943 &  1.544  &   0.641 \\
\hline
   & D-P & $-4.377$ & 4.027 &  1.130  &   0.648 \\
ND & D-D & $-4.518$ & 4.071 &  1.134  &   0.651 \\
   & G-G & $-4.553$ & 4.066 &  1.137  &   0.651 \\
   & I-I & $-4.553$ & 4.067 &  1.138  &   0.651 \\
\hline
   & D-P & $-3.853$ & 3.825 &  1.223  &   0.637 \\
NF & D-D & $-3.938$ & 3.851 &  1.226  &   0.639 \\
   & G-G & $-3.981$ & 3.849 &  1.236  &   0.639 \\
   & I-I & $-3.981$ & 3.849 &  1.236  &   0.639 \\
\hline
   & D-P & $-3.645$ & 3.805 &  1.380  &   0.635 \\
JA & D-D & $-3.710$ & 3.827 &  1.385  &   0.637 \\ 
   & G-G & $-3.762$ & 3.825 &  1.401  &   0.637 \\
   & I-I & $-3.762$ & 3.826 &  1.402  &   0.637 \\
\hline
   & D-P & $-3.460$ & 3.775 &  1.507  &   0.632 \\
JB & D-D & $-3.510$ & 3.792 &  1.511  &   0.633 \\
   & G-G & $-3.568$ & 3.793 &  1.535  &   0.634 \\
   & I-I & $-3.568$ & 3.794 &  1.535  &   0.634 \\
\hline
\end{tabular}
\end{center}
\end{table}

Table \ref{table11} summarizes the present results
with the MN force for the $\alpha \alpha$ RGM kernel.
The SB result shows the overbinding
of the $\hbox{}^9_{\Lambda}\hbox{Be}$ ground-state energy
by about 220 keV and too small excitation energy
of the $2^+$ excited state by about 130 keV.
Table \ref{table11} also shows a comparison with the results
by Hiyama et al.\ \cite{HI97} for the $G$-matrix based
effective $\Lambda N$ forces.
We find that their results are a little
lower than our results by about 70 - 90 keV.
Since their calculation is a variational calculation
using a smaller model space than ours,
this is not a convergence problem of the variational calculation.
A possible reason is the difference between OCM and RGM 
in the $\alpha \alpha$ part. They used $\alpha \alpha$ OCM,
while ours is $\alpha \alpha$ RGM.
The OCM usually gives more attractive results than the RGM.
In fact, it is well known that $3\alpha$ OCM usually gives
a larger binding energy than the $3\alpha$ RGM for
the ground state of the $3\alpha$ system \cite{SP80}.
A small difference in the exchange term of
the $\Lambda \alpha$ folding potential may also contribute
to this difference.

If we arrange the effective $\Lambda N$ forces
in Table \ref{table11} in the order of more attractive nature,
we find 
\begin{eqnarray}
& & {\rm ND}~(-7.483) > {\rm NF}~(-6.906) > {\rm SB}~(-6.837)
\nonumber \\
& & > {\rm NS}~(-6.742) > {\rm JA}~(-6.677) > {\rm JB}~(-6.474)\ .
\nonumber \\
\label{bel5}
\end{eqnarray}
The experimental value $-6.62 \pm 0.04~\hbox{MeV}$ is located
between JA and JB. However, this does not mean that the J{\"u}lich
potentials JA and JB are the most correct $\Lambda N$ interactions.
It is well known that the spin-spin central terms of these
J{\"u}lich potentials are completely wrong
and that they fail to reproduce the observed energy spectrum
of the $\hbox{}^4_{\Lambda}\hbox{H}$
and $\hbox{}^4_{\Lambda}\hbox{He}$ systems \cite{YA94}.
As for the $2^+$ excitation energy, all the results in 
Table \ref{table11} are between 2.90 - 2.93 MeV.
They are too small by 110 - 130 keV with respect to the average
value 3.04 MeV of the two resonances recently observed
by $\gamma$-ray spectroscopy \cite{AK02,TA03}.
Since the experimental error bars are
at most $\pm 40~\hbox{keV}$ even
in the $(K, \pi)$ reaction \cite{MA83},
this is a meaningful disagreement.
It would be interesting to examine the $\ell s$ splitting
of the $5/2^+$ - $3/2^+$ states,
by introducing a small $\Lambda N$ spin-orbit force
predicted by our quark-model interaction.
\begin{table}[t]
\caption{
Summary of the ground-state energy $E_{\rm gr}(0^+)$ and
the $2^+$ excitation energy $E_{\rm x}(2^+)$ in MeV,
calculated by solving the Faddeev equation
for the $\alpha \alpha \Lambda$ system in the $LS$ coupling scheme.
The $\alpha \alpha$ RGM kernel is generated
from the three-range MN force with $u=0.94687$
and $\nu=0.257~\hbox{fm}^{-2}$ for the h.o.\ width parameter
of the $\alpha$-clusters.
}
\label{table11}
\begin{center}
\renewcommand{\arraystretch}{1.2}
\setlength{\tabcolsep}{4mm}
\begin{tabular}{cccc}
\hline
$V_{\Lambda N}$ & \multicolumn{2}{c}{$E_{\rm gr}(0^+)$ (MeV)}
  & $E_{\rm x}(2^+)$ (MeV) \\
  & present & Ref.~\protect\cite{HI97} & \\
\hline
SB & $-6.837$  &  $-$    & 2.911 \\
NS & $-6.742$  & $-6.81$ & 2.912 \\
ND & $-7.483$  & $-7.57$ & 2.930 \\
NF & $-6.906$  & $-7.00$ & 2.925 \\
JA & $-6.677$  & $-6.76$ & 2.915 \\
JB & $-6.474$  & $-6.55$ & 2.907 \\
\hline
Exp't & \multicolumn{2}{c}{$-6.62 \pm 0.04$} & 3.024(3) \\
\protect\cite{AK02,TA03} & & & 3.067(3) \\
\hline
\end{tabular}
\end{center}
\end{table}

In order to show that the present $\alpha \alpha$ RGM kernel
gives a better result than simple $\alpha \alpha$ potentials,
we show in Table \ref{table12} some results
of $\alpha \alpha \Lambda$ Faddeev calculations
using the Ali-Bodmer potential, ABd \cite{AL66},
and the Buck, Friedrich, and Wheatley potential, BFW \cite{BF77}.
In these cases, there needs no self-consistent procedure
to determine $\varepsilon_{2\alpha}$.
We only use the SB potential for the $\Lambda \alpha$ interaction,
since results with other effective $\Lambda N$ forces
are easily evaluated from the above discussion
in the case of the $\alpha \alpha$ RGM kernel.
In these $\alpha$-particle models, we customarily use $\hbar^2/M_\alpha
=10.4465~\hbox{MeV}\cdot\hbox{fm}^2$ and $e^2=1.44
~\hbox{MeV}\cdot\hbox{fm}$.
The momentum discretization points with $n_1$-$n_2$-$n_3$
=15-10-5 are employed.
For the $\alpha \alpha$ Coulomb potential, the folding potential
of the cut-off Coulomb force with the $(0s)^4$ shell-model
wave function is used with $R_C=10~\hbox{fm}$.
The h.o.\ width parameter of the $(0s)^4$ $\alpha$-cluster
for this Gaussian folding
is $\nu=0.27127~\hbox{fm}^{-2}$ in the ABd case
and $\nu=0.257~\hbox{fm}^{-2}$ in the BFW case.
In the ABd case, this $\nu$ value corresponds
to the Coulomb-force parameter $\beta=\sqrt{3}/(2 \times 1.44)
=0.6014~\hbox{fm}^{-1}$ and the $\alpha$ rms
radius, $r_\alpha=(3/4\sqrt{\nu})=1.44~\hbox{fm}$.
Since this $\nu$ value is also used for the $\alpha$-cluster
folding for the $\Lambda N$ potential,
the $\Lambda \alpha$ bound-state
energy $E(\hbox{}^5_\Lambda \hbox{He})$ is
a little shifted from the fitted experimental
value $-3.12~\hbox{MeV}$.
[The different $\hbar^2/M_N$ value also affects this difference.]
Since the energy change is only about 0.06 MeV, we do not readjust
the potential parameters of the $\Lambda N$ force.
In the BFW case, the $\nu$ value, $0.257~\hbox{fm}^{-2}$,
corresponds to $\beta=\sqrt{4\nu/3}=0.58538~\hbox{fm}^{-1}$ and
the rms radius of the $\alpha$-cluster,
$r_\alpha=\sqrt{3}/(2\beta)=1.48~\hbox{fm}$.
In this case the difference of the $\Lambda N$ bound-state energy,
$0.054~\hbox{MeV}$, from $-3.12$ MeV is solely from
the different $\hbar^2/M_N$ value. 
The bound-state solutions of the BFW potential are used
for the pairwise Pauli-forbidden states. 
The elimination of the Pauli-forbidden components from the three-body
total wave function is always inspected by calculating
their squared norm, which is of the order of $10^{-13}$.

\begin{table}[b]
\caption{$\alpha \alpha \Lambda$ Faddeev calculations for
the $L^\pi=0^+$ ground state by
the Ali-Bodmer (ABd) \protect\cite{AL66} and Buck, Friedrich,
and Wheatley (BFW) \protect\cite{BF77} $\alpha \alpha$ potentials.
The SB $\Lambda N$ force is used
for the $\Lambda \alpha$ interaction.
The cut-off Coulomb force is included at the nucleon level
with $R_C=10~\hbox{fm}$.
The h.o.\ width parameters of the $\alpha$-clusters
are assumed to be $\nu=0.27127~\hbox{fm}^{-2}$ (ABd)
and $\nu=0.257~\hbox{fm}^{-2}$ (BFW).
The parameters $\hbar^2/M_\alpha=10.4465~\hbox{MeV}\cdot \hbox{fm}^2$
and $e^2=1.44~\hbox{MeV}\cdot \hbox{fm}$ are used.
Partial waves up to $\lambda_{\rm max}$ are included
in the $\alpha \alpha$-$\Lambda$ channel and those up
to ${\ell_1}_{\rm max}$ in the $\Lambda \alpha$-$\alpha$ channel.
The momentum discretization points with $n_1$-$n_2$-$n_3$
=15-10-5 are employed. 
The $\Lambda \alpha$ bound-state
energy $E(\hbox{}^5_\Lambda \hbox{He})$ for
the SB $\Lambda N$ force is given in the first column.
}
\label{table12}
\begin{center}
\renewcommand{\arraystretch}{1.0}
\setlength{\tabcolsep}{2mm}
\begin{tabular}{cccccc}
\hline
\multicolumn{6}{c}{ABd+SB} \\
\hline
$E(\hbox{}^5_\Lambda \hbox{He})$
 & $\lambda_{\rm max}$-${\ell_1}_{\rm max}$
 & $E(\hbox{}^9_\Lambda \hbox{Be})$
 & $\varepsilon_{2\alpha}$ & $\varepsilon_{\Lambda \alpha}$
 & $c_{(40)}$ \\
\hline
         & S-S & $-6.409$ & 0.970 &  $-0.503$ &  0.466 \\
         & D-P & $-7.091$ & 1.013 &  $-0.532$ &  0.497 \\
$-3.183$ & D-D & $-7.147$ & 1.013 &  $-0.526$ &  0.499 \\
         & G-G & $-7.153$ & 1.018 &  $-0.518$ &  0.498 \\
         & I-I & $-7.153$ & 1.018 &  $-0.517$ &  0.498 \\
\hline
\multicolumn{6}{c}{BFW+SB} \\
\hline
$E(\hbox{}^5_\Lambda \hbox{He})$
 & $\lambda_{\rm max}$-${\ell_1}_{\rm max}$
 & $E(\hbox{}^9_\Lambda \hbox{Be})$
 & $\varepsilon_{2\alpha}$ & $\varepsilon_{\Lambda \alpha}$
 & $c_{(40)}$ \\
\hline
         & S-S & $-5.544$ & 0.861 &  1.776 &  0.630 \\
         & D-P & $-6.971$ & 1.147 &  1.973 &  0.724 \\
$-3.066$ & D-D & $-7.038$ & 1.155 &  1.979 &  0.728 \\
         & G-G & $-7.043$ & 1.161 &  1.979 &  0.728 \\
         & I-I & $-7.043$ & 1.161 &  1.980 &  0.728 \\
\hline
\end{tabular}
\end{center}
\end{table}

We find that the $\alpha \alpha \Lambda$ ground-state energy
by the ABd potential is lower than the result of
the MN force in \eq{bel1} by 0.3 MeV.
Note that even in this case the energy gain from the higher partial
waves than the $S$ wave is appreciable, i.e., 0.7 MeV.
This implies that the $S$-wave assumption
adopted by Filikhin and Gal \cite{FI02} is not valid. 
They used a little different version of the Ali-Bodmer
potential (type (a) with 125 MeV modified by 120 MeV) and
obtained $E(\hbox{}^9_\Lambda\hbox{Be})=-6.55~\hbox{MeV}$ in
the $S$-wave approximation.
We expect an energy gain of about 0.7 MeV from the higher partial waves
and their result is overbound,
in comparison with the experimental value, $-6.62 \pm 0.04~\hbox{MeV}$.
In Table \ref{table12}, we find that the BFW potential
gives a better result than the Ali-Bodmer force,
but the energy is still lower than in the MN force case
by 0.2 MeV. In this case we find that the effect of 
partial waves higher than the $S$ wave
is quite appreciable, i.e., $-1.5$ MeV.
This is of course due to the inner oscillation of the relative
wave function between the two $\alpha$-clusters
in the $\alpha \alpha \Lambda$ ground state. 
The shell-model like $(40)$ components are about 0.7
in amplitude, which is appreciably larger
than $c_{(40)} \sim 0.5$ in the Ali-Bodmer case.

\section{Summary}

The three-cluster Faddeev formalism using two-cluster
resonating-group method (RGM) kernels
opens a way to solve few-baryon systems interacting via
quark-model baryon-baryon interactions without spoiling
essential features of the RGM kernel;
i.e., the non-locality, the energy dependence proportional
to the exchange normalization kernel,
and the existence of pairwise Pauli-forbidden states
in some specific channels. In this paper, we have applied
this formalism to three-cluster systems
involving $\alpha$-clusters;
i.e., the $3\alpha$ and $\alpha \alpha \Lambda$ systems.
These systems involve all of the above three features
for the microscopic interactions between
composite particles. In particular, the $\alpha \alpha$ interaction
is a prototype of composite-particle interactions, in which the
fully microscopic RGM calculation is easy and very successful.
It, however, involves a somewhat complex kernel structure
composed of three non-trivial Pauli-forbidden states, and
the energy-dependence of the interaction is rather strong
in the Pauli-allowed model space.
In the present Faddeev formulation, the Pauli-forbidden
components between pairwise clusters are completely
eliminated from the total wave function of the three clusters.
This can be achieved by introducing a special type
of RGM $\widetilde{T}$-matrix calculated from
the two-cluster RGM kernel, which satisfies the $T$-matrix
version of the orthogonality conditions to the relative
motion between two clusters. The on-shell and half off-shell
properties of the $\widetilde{T}$-matrix are just the same as
those of the ordinary $T$-matrix. 
This RGM $\widetilde{T}$-matrix involves a relative energy
of two clusters as a parameter, which is determined
self-consistently by calculating the expectation value
of the two-cluster Hamiltonian with respect to the 
total wave function resulting from the Faddeev equation.
The Faddeev equation using $\widetilde{T}$-matrices is
equivalent to the pairwise orthogonality condition model (OCM)
of three-cluster systems, interacting via two-cluster RGM kernels. 
A nice point of this formalism is that the
underlying nucleon-nucleon ($NN$) and
hyperon-nucleon ($YN$) interactions are more directly related
to the structure of three-cluster systems
than in the models assuming simple two-cluster potentials.

We have first applied the present formalism
to the ground state of the $3\alpha$ system
by using three different types of effective $NN$ forces,
the two-range Volkov forces, No.\,1 (VN1) and No.\,2 (VN2),
and the three-range Minnesota (MN) force.
The three-range MN force reproduces
the $S$-, $D$- and $G$-wave $\alpha \alpha$ phase shifts
quite well in the simple $(0s)^4$-model of the $\alpha$ clusters.
The comparison with the $3\alpha$ RGM calculation has shown that
the present three-cluster formalism using
only the $\alpha \alpha$ RGM kernel
gives a good approximation to the microscopic $3\alpha$ model.
The difference of the ground-state 
energies predicted by these two models is less than 2 MeV.
The effect of the antisymmetrization among
three $\alpha$-clusters, which is neglected in our formulation,
is attractive and is not so large, as long as the Pauli-allowed
model space of the $3\alpha$ system is properly treated.
It is also shown that the three-range MN force gives
a lower ground-state energy than the two-range VN1 and VN2 forces, 
resulting in a somewhat large overbinding of 2 - 4 MeV,
if the $3\alpha$ ground-state energy
is measured from the $3\alpha$ threshold.

The application to the $\hbox{}^9_{\Lambda}\hbox{Be}$
system has proved that our three-cluster formalism is soundly
extended to the systems with two identical clusters,
in addition to the systems of three identical clusters
like the $3\alpha$ system and the triton system.
Here we have introduced a new effective $\Lambda N$ force,
called the SB force, which is made from
the quark-model predictions of the $\Lambda N$ phase shifts
by using an inversion method based on supersymmetric
quantum mechanics \cite{SB97}.
The SB force consists of two simple two-range Gaussian potentials
which reproduce the low-energy behavior
of the $\hbox{}^1S_0$ and $\hbox{}^3S_1$ $\Lambda N$ phase shifts
predicted by $\Lambda N$-$\Sigma N$ coupled-channel
RGM calculations using the model fss2 \cite{B8B8}.
Since any central and single-channel effective $\Lambda N$ force 
leads to the well-known overbinding problem
of $\hbox{}^5_{\Lambda}\hbox{He}$ by about 2 MeV \cite{KO03},
the attractive part of the $\hbox{}^3S_1$ $\Lambda N$ potential
is reduced by about $10\,\%$ to reproduce
the empirical $\Lambda$-separation energy, $B^{\rm exp}_\Lambda
(\hbox{}^5_\Lambda \hbox{He})=3.12\pm 0.02$ MeV.
The odd-state $\Lambda N$ force is assumed to be zero
(pure Serber type). In addition to this SB force,
we have also used the effective $\Lambda N$ forces
in Ref.~\cite{HI97} for comparison.
The $\Lambda \alpha$ interactions are generated from
these $\Lambda N$ effective forces by the folding procedure
with respect to the $(0s)^4$ h.o.\ wave function
of the $\alpha$ clusters.

In the $\alpha \alpha \Lambda$ Faddeev calculation,
enough partial waves up to $\lambda_{\rm Max}={\ell_1}_{\rm Max}=6$
are included both in the $\alpha \alpha$ and $\Lambda \alpha$ pairs
since the relative wave functions between two $\alpha$-clusters
are oscillating at least in the relative $S$- and $D$-waves.
The detailed analysis shows that the partial waves up to
the $D$-wave are sufficient if we do not mind a 10 keV inaccuracy.
If we wish to obtain a 1 keV accuracy, we need to take into account
at least up to the $G$-wave. This implies that the partial wave
truncation is very efficient even in the present Faddeev formalism.
The energy gain due to partial waves higher than the $S$-wave
is about 1 MeV for the VN2 force and 1.2 MeV for the MN force,
when these $\alpha \alpha$ interactions are used
in combination with the SB force
for the $\Lambda \alpha$ interaction.
The Coulomb effect between the two $\alpha$-clusters is
included by a cut-off Coulomb force at the nucleon level.
The cut-off radius, $R_C=10$ - 14 fm seems to be sufficient
for a 1 - 2 keV accuracy. In the present formalism,
the structure change of two $\alpha$-clusters inside
$\hbox{}^9_{\Lambda}\hbox{Be}$ is clearly identified
by calculating the kinetic-energy contribution
in the two-$\alpha$ expectation value $\varepsilon_{2\alpha}$.
The comparison of the Coulomb contributions in the $\alpha \alpha$ 
bound state, $\varepsilon_{2\alpha}$
and the $\hbox{}^9_{\Lambda}\hbox{Be}$ ground state with
respect to the change of $R_C$ is very useful to measure
the compactness of the two-$\alpha$ configurations in various
environments. It is confirmed that the $0^+$ ground state
and the $2^+$ exited state of $\hbox{}^9_{\Lambda}\hbox{Be}$ are
well described by the contracted two-$\alpha$ cluster structure with
a weakly coupled $\Lambda$-particle in the dominant $S$-wave
component.
In the present calculation using only central forces,
the three-range MN force and the SB potential with the pure-Serber
character can reproduce the ground-state and excitation energies
of $\hbox{}^9_{\Lambda}\hbox{Be}$ within an accuracy
of 100 - 200 keV. The results in Ref.~\cite{HI97} based
on the OCM framework are also confirmed within 100 keV accuracy.
On the other hand, the simple $\alpha$-particle
model using the Ali-Bodmer $\alpha \alpha$ potential,
ABd \cite{AL66}, and the OCM using the deep Buck, Friedrich,
and Wheatley $\alpha \alpha$ potential, BFW \cite{BF77}, with
bound-state Pauli-forbidden states give an overbinding
of the $\hbox{}^9_{\Lambda}\hbox{Be}$ ground state
by 530 keV and 420 keV, respectively,
when the SB force is used for the $\Lambda \alpha$ interaction.
Although these energies are rather similar, the effect of 
partial waves higher than the $S$-wave is very different; i.e.,
0.7 MeV in the Ali-Bodmer case and 1.5 MeV in the BFW case.
It is natural that the $\alpha \alpha$ interactions
which yield an oscillatory behavior
of the $\alpha \alpha$ relative wave functions, like our
RGM kernel and the BFW potential, need more 
partial waves with a larger energy gain.

There are still many problems left for the future studies.
First of all, the readjustment of the $\hbox{}^3S$ attractive
part of the SB $\Lambda N$ potential is unsatisfactory
from the viewpoint of using the fundamental baryon-baryon
interactions. The Brueckner rearrangement effect
in $\hbox{}^5_\Lambda \hbox{He}$ is fairly large even
for the rather stable $\alpha$-cluster \cite{KO03}.
In this sense, there is still no consistent description
of the $s$-shell and $p$-shell hypernuclei even at the
level of using effective baryon-baryon interactions.
A microscopic description of the $\Lambda \alpha$ interaction
may need a more detailed analysis based on the $G$-matrix
theory, for which the folding formula given in Appendix B is
very useful. In order to describe
the $\hbox{}^9_{\Lambda}\hbox{Be}$ excited states realistically,
we need to introduce the $\Lambda \alpha$ spin-orbit force
and solve the Faddeev equation in the $jj$-coupling scheme.
The recent $\gamma$-ray spectroscopy experiment \cite{AK02,TA03} 
indicates a very small spin-orbit splitting for the
possible $5/2^+$ and $3/2^+$ resonances.
It is interesting to examine the $LS$ components of
the quark-model $\Lambda N$ interaction, in which the
antisymmetric $LS$ interaction ($LS^{(-)}$) is
by about a factor two larger than in the Nijmegen models.
We expect a large cancellation between
the ordinary $LS$ interaction and this $LS^{(-)}$ interaction.
An interesting application of the present Faddeev formalism
and the $\Lambda \alpha$ $T$-matrix
derived in this study is to the recent Nagara
event \cite{Nagara} for $\hbox{}^{\ 6}_{\Lambda \Lambda}\hbox{He}$.
For the $\Lambda \Lambda$ interaction, we can use
the coupled-channel $\Lambda \Lambda$-$\Xi N$-$\Sigma \Sigma$
$\widetilde{T}$-matrix of the quark-model interaction, fss2.
A preliminary result \cite{HE6LL} shows that fss2 is at present the only model
which can reproduce an appropriate strength
of the $\Lambda \Lambda$ interaction,
$\Delta B^{\rm exp}_{\Lambda \Lambda}=1.01
\pm 0.20$ MeV, deduced from the Nagara event.
In a separate paper \cite{hypt},
we will also report another application
of the present three-cluster Faddeev formalism
to the hypertriton system,
in which the quark-model $NN$ and $YN$ interactions are
explicitly used in the $\Lambda NN$ and $\Sigma NN$ coupled-channel
Faddeev formalism. In this system, a complete Pauli-forbidden
state at the quark level exists
in the $\Lambda N$-$\Sigma N$ subsystem. 

\begin{acknowledgments}
This work was supported by Grants-in-Aid for Scientific
Research (C) from the Japan Society for the Promotion
of Science (JSPS) (Nos.~15540270, 15540284, and 15540292).
Y. Fujiwara wishes to thank the FNRS foundation
of Belgium for making his visit to the Free University of Brussels
possible during the summer, 2002.
\end{acknowledgments}

\appendix

\section{Rearrangement factors of three-cluster systems
with two identical particles}

In this appendix, we give a brief comment on the definition
of the rearrangement factors in the Dirac notation
for general three-body systems
with two identical particles or clusters.
The incorporation of spin-isospin degrees of freedom is
essential for further applications to the hypertriton system \cite{hypt}
and the $\Lambda \Lambda \alpha$ system \cite{HE6LL}.
When one uses the Dirac notation, it is important to fix
a coordinate system of the representation.
We choose the standard system
of the Jacobi coordinates with $\gamma=3$, and
introduce the Jacobi coordinates in the momentum
space, $\bp=\bp_3$ and $\bq=\bq_3$.
The other Jacobi coordinates $\bp_1$, $\bq_1$, etc. are
similarly defined.
For an arbitrary function $\psi(\bp, \bq; 123)$ in $\gamma=3$,
the effect of the cyclic permutation $P_{(123)}$ of the
symmetric group $S_3$ is
\begin{eqnarray}
P_{(123)} \psi(\bp_2, \bq_2; 312)
& = & {P_{(123)}}^2 \psi(\bp_1, \bq_1; 231)
\nonumber \\
& = & \psi(\bp_3, \bq_3; 123)\ ,
\label{a1}
\end{eqnarray}
where 123 in $\psi(\bp, \bq; 123)$ stands for the spin-isospin
variables.
For the transposition $P_{(12)}$, \eq{form1} yields
\begin{eqnarray}
& & P_{(12)} \psi(\bp_3, \bq_3; 123)
=\psi(-\bp_3, \bq_3; 213)\ , \nonumber \\
& & P_{(12)} \psi(\bp_1, \bq_1; 231)
=\psi(-\bp_2, \bq_2; 132)\ , \nonumber \\
& & P_{(12)} \psi(\bp_2, \bq_2; 312)
=\psi(-\bp_1, \bq_1; 321)\ .
\label{a2}
\end{eqnarray}
Note that the momentum suffix $\alpha$ in $\bp_\alpha$,
$\bq_\alpha$, and the sign of $\bp_\alpha$, etc., are
uniquely specified by the sequence of 123.
For example, $\langle \widehat{\bp}_2, \widehat{\bq}_2
|\beta \rangle$ in \eq{form8} actually
implies $\langle \widehat{\bp}_2, \widehat{\bq}_2;
312|\beta \rangle$.
In the following, we always use an abbreviated
notation, $\psi=\psi(\bp_3, \bq_3; 123)$, in the standard
coordinate system $\gamma=3$. The total wave function,
$\Psi(\bq_3, \bq_3; 123)$, in \eq{form3} is then compactly
expressed as
\begin{eqnarray}
\Psi=\psi-P_{(12)}{P_{(123)}}^2\varphi+{P_{(123)}}^2\varphi\ .
\label{a4}
\end{eqnarray}
If we write the Faddeev equation in terms of $\psi$ and $\varphi$,
it reads
\begin{eqnarray}
\psi & = & G_0 \widetilde{T} \left(1 \pm P_{(12)} \right)
{P_{(123)}}^2 \varphi\ , \nonumber \\
\varphi & = & G_0 T \left[ P_{(123)} \psi
\pm P_{(23)} \varphi \right]\ ,
\label{a7}
\end{eqnarray}
with $\widetilde{T}=T_{12}$ and $T=P_{(123)}T_{13}
{P_{(123)}}^{-1}$, where $T_{12}$ and $T_{13}$ are
the two-body $T$-matrices in the three-body space.

The definition of the rearrangement factors
in the Dirac notation is based on the assumption
\begin{eqnarray}
& & \langle \bp_3, \bq_3; 123|{P_{(123)}}^2 \varphi \rangle
={P_{(123)}}^2 \varphi(\bp_3, \bq_3; 123)
\nonumber \\
& & =\varphi(\bp_2, \bq_2; 312)
=\int d \bp^\prime d \bq^\prime~\delta(\bp^\prime-\bp_2)
~\delta(\bq^\prime-\bq_2)
\nonumber \\
& & \times {P^{(\sigma \tau)}_{(123)}}^2
\varphi(\bp^\prime, \bq^\prime; 123)\ ,
\label{a8}
\end{eqnarray}
where the function $\varphi(\bp_3, \bq_3; 123)$ is in \eq{a4}
and $P^{(\sigma \tau)}_{(123)}$ operates only on the spin-isospin
variables of $\varphi(\bp^\prime, \bq^\prime; 123)$.
With this $\varphi$ in the $\beta=2$ channel in mind,
the standard procedure of the partial wave decomposition
gives the following definition for the first-type
rearrangement factor $g_{\gamma \beta}(q, q^\prime; x)$:
\begin{widetext}
\begin{eqnarray}
& & \langle p, q, \gamma |{P_{(123)}}^2| p^\prime, q^\prime, \beta
\rangle_{3-2}
={1 \over 2} \int^1_{-1} dx {\delta(p-p_1) \over p^{\lambda+2}}
{\delta(p^\prime-p_2) \over {p^\prime}^{\ell_1+2}}
g_{\gamma \beta}(q, q^\prime; x)
= \sum_{123} \int d \widehat{\bp} d \widehat{\bq}
d \widehat{\bp}^\prime d \widehat{\bq}^\prime
~\langle \gamma |
\widehat{\bp}, \widehat{\bq}; 123 \rangle
\nonumber \\
& & \times
~\delta \left(\bp+\bq^\prime
+{m_2 \over m_2+m_1}\bq\right)
\delta \left(\bp^\prime-\bq-{m_3
\over m_3+m_1}\bq^\prime\right)
~{P^{(\sigma \tau)}_{(123)}}^2
~\langle \widehat{\bp}^\prime, \widehat{\bq}^\prime;
123|\beta \rangle \ .
\label{a14}
\end{eqnarray}
Here, $p_1$ and $p_2$ are given in \eq{form11c} with
a general mass factor, $\zeta=(4m_3/m_1)$.
With this mass modification, \eq{form12} is valid with
a more complete reduced rearrangement factor
\begin{eqnarray}
g^{\lambda_1 \lambda^\prime_1 k}_{\gamma \beta}
=\left\{ \begin{array}{ll}
(-1)^\lambda
~G^{\lambda_1 \lambda^\prime_1 k L}_{(\lambda \ell),
(\ell_1 \ell_2)}
~\langle STT_z; \gamma ||{P^{(\sigma \tau)}_{(123)}}^2||STT_z;
\beta \rangle
& (LS{\rm -coupling}) \\
(-1)^\lambda
\sum_{LS} \left[\begin{array}{ccc}
\lambda & s_1 & I \\
\ell    & s_2 & j \\
L       & S   & J   \\
\end{array} \right]
\left[\begin{array}{ccc}
\ell_1 & s^\prime_1 & j_1 \\
\ell_2 & s^\prime_2 & j_2 \\
L      & S          & J   \\
\end{array} \right]
G^{\lambda_1 \lambda^\prime_1 k L}_{(\lambda \ell),(\ell_1 \ell_2)}
~\langle STT_z; \gamma ||{P^{(\sigma \tau)}_{(123)}}^2||STT_z;
\beta \rangle
& (jj{\rm -coupling}) \\
\end{array} \right. .
\label{a18}
\end{eqnarray}
Here the square bracket implies the unitary form
of the $9j$ coefficients and
the quantum numbers are specified by
\begin{eqnarray}
\left\{
\begin{array}{lll}
|\gamma \rangle = |[(\lambda \ell) LS] JJ_z;TT_z \rangle
& |\beta \rangle=|[(\ell_1 \ell_2) LS] JJ_z; TT_z \rangle
& (LS{\rm -coupling}) \\
|\gamma \rangle = |[(\lambda s_1)I (\ell s_2)j] JJ_z; TT_z
& |\beta \rangle=|[(\ell_1 s^\prime_1)j_1
(\ell_2 s^\prime_2)j_2] JJ_z; TT_z \rangle
& (jj{\rm -coupling}) \\
\end{array} \right. \ .
\label{a19}
\end{eqnarray}
The angular-momentum
factors $G^{\lambda_1 \lambda^\prime_1 k L}_{(\lambda \ell),
(\ell_1 \ell_2)}$ with $\lambda_1=0 \sim \lambda$,
$\lambda^\prime_1=0 \sim \ell_1$ are given by
\begin{eqnarray}
& & G^{\lambda_1 \lambda^\prime_1 k L}_{(\lambda \ell),
(\ell_1 \ell_2)}
=G^{\lambda^\prime_1 \lambda_1 k L}_{(\ell_1
\ell_2), (\lambda \ell)}
=\left[{(2\lambda+1)! (2\ell_1+1)! \over
(2\lambda_1)! (2\lambda_2)! (2\lambda^\prime_1)!
(2\lambda^\prime_2)!}\right]^{1 \over 2}
\widehat{\lambda}\widehat{\ell}\widehat{\ell_1}
\widehat{\ell_2}
\nonumber \\
& & \times
\sum_{f, f^\prime} \langle \lambda_2 0 \ell 0 | f 0 \rangle
\langle \lambda^\prime_2 0 \ell_2 0 | f^\prime 0 \rangle
\langle k 0 \lambda_1 0 | f^\prime 0 \rangle
\langle k 0 \lambda^\prime_1 0 | f 0 \rangle
\left\{ \begin{array}{ccc}
f         &    L      & \lambda_1 \\
\lambda   & \lambda_2 &    \ell    \\
\end{array} \right\}
\left\{ \begin{array}{ccc}
f^\prime  &    L      & \lambda^\prime_1 \\
\ell_1 & \lambda^\prime_2 & \ell_2 \\
\end{array} \right\}
\left\{ \begin{array}{ccc}
\lambda^\prime_1 & f^\prime & L \\
\lambda_1        &    f     & k \\
\end{array} \right\}\ ,
\label{a17}
\end{eqnarray}
where $\widehat{\lambda}=\sqrt{2\lambda+1}$ etc. and
$\lambda_2=\lambda-\lambda_1$, $\lambda^\prime_2=\ell_1
-\lambda^\prime_1$.
In the spin-isospin reduced matrix elements of \eq{a18}, the permutation
operator ${P^{(\sigma \tau)}_{(123)}}^2$  does
not change the total spin and isospin values, $S$ and $TT_z$.

The other types of rearrangement factors are obtained
in a similar way. First, the symmetry of the matrix elements
yields
\begin{eqnarray}
\langle p, q, \beta |P_{(123)}|
p^\prime, q^\prime, \gamma \rangle_{2-3}
=\langle p^\prime, q^\prime, \gamma |{P_{(123)}}^2|
p, q, \beta \rangle_{3-2}\ .
\label{a20}
\end{eqnarray}
The rearrangement factor for the matrix
element $\langle \varphi|P_{(23))}|\varphi \rangle$ needs
a little care, since the mass assignment of the three particles
is made in the standard Jacobi coordinates $\gamma=3$.
We first use $P_{(23)}=P_{(123)}P_{(12)}{P_{(123)}}^{-1}$ and
write the matrix element as
\begin{eqnarray}
\langle \varphi|P_{(23)}|\varphi \rangle
=\sum_{123} \int d \bp_3 d \bq_3~\varphi^*(\bp_2, \bq_2; 312)
~\varphi(-\bp_1, \bq_1; 321)\ .
\label{a21}
\end{eqnarray}
The corresponding rearrangement factor in the Dirac notation is
given by
\begin{eqnarray}
\langle p, q, \beta |P_{(23)}| p^\prime, q^\prime,
\beta^\prime \rangle_{2-1} 
& = & \sum_{123} \int d \widehat{\bp} d \widehat{\bq}
d \widehat{\bp}^\prime
d \widehat{\bq}^\prime
~\langle \beta |
\widehat{\bp}, \widehat{\bq}; 123 \rangle
~\delta \left(\bp+\bq^\prime+{m_1 \over m_1+m_3}\bq\right)
~\delta \left(\bp^\prime+\bq+{m_2 \over m_2+m_3}\bq^\prime\right)
\nonumber \\
& & \times~P^{(\sigma \tau)}_{(23)}
~\langle \widehat{\bp}^\prime, \widehat{\bq}^\prime;
123|\beta^\prime \rangle\ ,
\label{a22}
\end{eqnarray}
from which the results
in Eqs.~(\ref{form14}) and (\ref{form15}) are easily obtained.
Note that the rearrangement factor \eq{a22} is symmetric
with respect to the interchange between $p,~q,~\beta$ and
$p^\prime,~q^\prime,~\beta^\prime$, since $m_1=m_2$.
\end{widetext}

\section{A useful formula for the \mib{\Lambda \alpha} Born kernel}

The general procedure to calculate Born kernels of the $s$-shell
clusters, developed in Ref.~\cite{LSRGM}, can also be used
to calculate the $\Lambda \alpha$ Born kernel
\begin{eqnarray}
& & V(\bq_f, \bq_i)=\langle e^{i\bq_f \br}|V|e^{i\bq_i \br} \rangle
\nonumber \\
& & =\langle e^{i\bq_f \br} \xi_\Lambda\phi_\alpha|
\sum^5_{j=2} v_{1j} |e^{i\bq_i \br} \xi_\Lambda \phi_\alpha
\rangle\ ,
\label{b1}
\end{eqnarray}
where $\phi_\alpha$ is the internal wave function
of the $\alpha$ cluster, $\xi_\Lambda$ is the spin wave function
of the $\Lambda$ particle and $v_{1j}$ is
an effective $\Lambda N$ interaction. The essential part of
this method lies in the correct treatment of the c.m. motion
which is handled by the procedure given in Ref.~\cite{NA95}.
This method makes it possible to deal with the most general form
of the $\Lambda N$ interaction with non-static effects
like the $G$-matrix $\Lambda N$ interaction.
In this method, $V(\bq_f, \bq_i)$ in \eq{b1} is calculated from
an integral form of the GCM kernel through 
\begin{eqnarray}
& & V(\bq_f, \bq_i)=\langle \delta(\bX_G) e^{i\bq_f \br}
\xi_\Lambda \phi_\alpha |\sum^5_{j=2} v_{1j}|
1 \cdot e^{i\bq_i \br} \xi_\Lambda \phi_\alpha \rangle \nonumber \\
& & =\left({\gamma \over 2\pi}\right)^{3 \over 2}
e^{{1 \over 4\gamma}({\bq_f}^2+{\bq_i}^2)}
\int d \ba\,d \bb~e^{-i\bq_f \cdot \ba+i \bq_i \cdot \bb}
G(\ba, \bb) ,
\nonumber \\
\label{b2}
\end{eqnarray}
with
\begin{eqnarray}
G(\ba, \bb) & = & \left({\gamma_G \over 2\pi}\right)^{3 \over 2}
\int d \bR~\langle \psi_\Lambda(\ba) \psi_\alpha(0)|
\nonumber \\
& & \times \sum^5_{j=2} v_{1j}\,|\psi_\Lambda(\bR+\bb) \psi_\alpha(\bR)
\rangle\ . \label{b3}
\end{eqnarray}
Here, $\gamma_G=(4+\zeta)\nu$, $\gamma=4\zeta\nu/(4+\zeta)$
with $\zeta=M_\Lambda/M_N$, and $\psi_\Lambda(\bR)$
and $\psi_\alpha (\bR)$ are the h.o.\ shell model
wave functions of $\Lambda$ and $\alpha$, centered at $\bR$,
with the width parameters $\zeta\nu$ and $\nu$, respectively. 
First we calculate spatial integrals for the spatial
part $u$ in $v_{1j}=u_{1j} \omega_{1j}$. These four integrals
with $j=2$ - 5 are all equal because of the antisymmetric
property of the $\alpha$ cluster. We need to calculate
spatial integrals for $u=u(\bx_1-\bx_2)$ and $u=u(\bx_1-\bx_2)P_r$,
which we call the direct term and the exchange term, respectively.
It is important to note that the space exchange operator, $P_r$,
operates only on the single-particle coordinates $\bx_1$
and $\bx_2$, and {\em does not} exchange the $\Lambda$ and $N$
masses. The procedure to interchange these masses $M_\Lambda$
and $M_N$ simultaneously like in Ref.~\cite{HI97} leads
to an erroneous expression [see Eq.\,(A.1) of \cite{HI97}],
which is apparently wrong since
the RGM kernel $\langle \delta(\br-\ba)\xi_\Lambda \phi_\alpha
|\sum^5_{j=2} v_{1j}|\delta(\br-\bb)
\xi_\Lambda \phi_\alpha \rangle$ should not involve the
mass dependence. The correct expression is the one in which one
sets $M_N=M_\Lambda$ in their Eq.\,(A.1) [see \eq{b10} below].
The most general form of the two-body $\Lambda N$ matrix elements
for the translationally invariant $u$ is parameterized as
\begin{eqnarray}
\langle \bp_1 \bp_2|u|{\bp_1}^\prime {\bp_2}^\prime \rangle
={1 \over (2\pi)^3} \delta(\bP-\bP^\prime)
u(\bk^\prime, \bq^\prime; \bP)\ ,
\label{b4}
\end{eqnarray}
with $\bp=(\bp_1-\zeta \bp_2)/(\zeta+1)$,
$\bP=\bp_1+\bp_2$ (also $\bp^\prime$, $\bP^\prime$ for ${\bp_1}^\prime$,
${\bp_2}^\prime$), and $\bk^\prime=\bp-\bp^\prime$,
$\bq^\prime=(\bp+\bp^\prime)/2$.
For the matrix element \eq{b4}, the spatial part of the GCM kernel
in \eq{b3} is calculated to be
\begin{eqnarray}
& & G^{\rm space}(\ba, \bb) \nonumber \\
& & ={1 \over (2\pi)^6 \nu^3}
\left({4+\zeta \over 3\zeta}\right)^{3 \over 2}
\int d \bP\,d \bk^\prime\,d \bq^\prime
~u(\bk^\prime, \bq^\prime; \bP) \nonumber \\
& & \times \exp \left\{
-{1 \over 6\nu}{\zeta+4 \over \zeta+1}\bP^2
-{1 \over 2\nu}{\zeta+1 \over \zeta}\left({\bq^\prime}^2
+{1 \over 4}{\bk^\prime}^2\right) \right. \nonumber \\
& & \left. +i(\ba-\bb)\cdot \left(\bq^\prime
+{\zeta \over \zeta+1}\bP\right)
+i{1 \over 2}(\ba+\bb)\cdot \bk^\prime\right\}\ .\nonumber \\
\label{b7}
\end{eqnarray}
If we use \eq{b7} in \eq{b2}, we can perform the integrals 
over $\ba$ and $\bb$ and obtain two delta functions.
Thus we can perform the integrals
over $\bk^\prime$ and $\bq^\prime$ and obtain a compact formula
\begin{eqnarray}
& & V^{\rm space}(\bq_f, \bq_i)=e^{-{3 \over 32\nu}\bk^2}
\left({2 \over 3\pi \nu}\right)^{3 \over 2}
\int d \bP~e^{-{2 \over 3\nu} \bP^2}
\nonumber \\
& & \times u\left(\bk, {\zeta+4 \over 4(\zeta+1)}\bq
-{\zeta \over \zeta+1}\bP; \bP+{3 \over 4} \bq \right)\ ,
\label{b8}
\end{eqnarray}
where $\bk=\bq_f-\bq_i$ and $\bq=\left(\bq_f+\bq_i\right)/2$.

For a simple local Gaussian interaction, we find
\begin{eqnarray}
& & u(\bk, \bq; \bP)=\left({\pi \over \kappa}\right)^{3 \over 2}
\exp \left\{ -{\bk^2 \over 4\kappa}\right\}
\quad {\rm for}~u(r)=e^{-\kappa r^2},\nonumber \\
& & u(\bk, \bq; \bP)=\left({\pi \over \kappa}\right)^{3 \over 2}
\exp \left\{ -{1 \over \kappa}\left(\bq
+{1 \over 2}{\zeta-1 \over \zeta+1}
\bP\right)^2 \right\} \nonumber \\
& & \hspace{30mm} {\rm for} \quad u(r)=e^{-\kappa r^2}P_r\ .
\label{b9}
\end{eqnarray}
Then the $\bP$ integral is carried out easily and we obtain
\begin{eqnarray}
& & V_d(\qf, \qi)=\left({\pi \over \kappa}\right)^{3 \over 2}
\exp \left\{ -{1 \over 4}\left({3 \over 8\nu}
+{1 \over \kappa}\right)\bk^2 \right\}\nonumber \\
& & \qquad {\rm for} \quad u(r)=e^{-\kappa r^2}\ ,
\nonumber \\
& & V_e(\qf, \qi)=\left({8\pi \over 3}
{1 \over \nu+{8 \over 3}\kappa}\right)^{3 \over 2}
\exp \left\{ -{3 \over 32\nu}\bk^2 \right. \nonumber \\
& & \left. -{25 \over 24}{1 \over \nu
+{8 \over 3}\kappa}\bq^2 \right\}
\quad {\rm for} \quad u(r)=e^{-\kappa r^2} P_r\ .
\label{b10}
\end{eqnarray}
If we further incorporate the spin-isospin factors,
the full $V(\qf, \qi)$ is given by
\begin{eqnarray}
V(\qf, \qi)=X_d~V_d(\qf, \qi)+X_e~V_e(\qf, \qi)\ ,
\label{b11}
\end{eqnarray}
with the spin-isospin factors defined by
\begin{eqnarray}
\left\{ \begin{array}{c}
X_d \\
X_e \\
\end{array}\right\}
=\langle \xi_\Lambda \chi_\alpha |
\sum^5_{j=2} \left\{ \begin{array}{c}
\omega^d_{1j} \\
\omega^e_{1j} \\
\end{array} \right\} |\xi_\Lambda \chi_\alpha \rangle\ .
\label{b12}
\end{eqnarray}
Here $\chi_\alpha$ is the spin-isospin wave function
of the $\alpha$-cluster.
The partial wave decomposition of \eq{b10} is given by
\begin{widetext}
\begin{eqnarray}
V^d_\ell(q_f, q_i; \kappa) & = &
\left({\pi \over \kappa}\right)^{3 \over 2}
\exp \left\{ -{1 \over 4}\left({3 \over 8\nu}+{1 \over \kappa}\right)
\left({q_f}^2+{q_i}^2\right) \right\}
~i_\ell \left({1 \over 2}\left({3 \over 8\nu}+{1 \over \kappa}\right)
q_f q_i \right)\ ,
\nonumber \\
V^e_\ell(q_f, q_i; \kappa) & = & \left({8\pi \over 3}
{1 \over \nu+{8 \over 3}\kappa}\right)^{3 \over 2}
\exp \left\{ -{1 \over 4} \left({3 \over 8\nu}
+{25 \over 24}{1 \over \nu+{8 \over 3}\kappa}\right)
\left({q_f}^2+{q_i}^2\right) \right\}
~i_\ell \left({1 \over 2}\left({3 \over 8\nu}
-{25 \over 24}{1 \over \nu+{8 \over 3}\kappa}\right)
q_f q_i \right)\ ,\nonumber \\
\label{b13}
\end{eqnarray}
where $i_\ell(x)=i^\ell j_\ell(-ix)$ is the spherical Bessel
function of imaginary argument.
\end{widetext}

\end{document}